# Visualizing hot carrier dynamics by nonlinear optical microscopy at the atomic length scale

Yang Luo[1,*], Shaoxiang Sheng[1,*], Andrea Schirato[2,3,*], Alberto Martin-Jimenez[1,4], Giuseppe Della Valle[2,5,+], Giulio Cerullo[2,5], Klaus Kern[1,6], Manish Garg[1,+]

[1] Max Planck Institute for Solid State Research, Heisenbergstr. 1, 70569 Stuttgart, Germany

[2] Dipartimento di Fisica,Politecnico di Milano, Piazza L. da Vinci 32, 20133 Milano, Italy

[3] Department of Physics and Astronomy, Rice University, 6100 Main St Houston, Texas 77005, United States

[4] Instituto Madrileño de Estudios Avanzados en Nanociencia (IMDEA Nanociencia), Calle Faraday 9, 28049 Madrid, Spain

[5] Istituto di Fotonica e Nanotecnologie – Consiglio Nazionale delle Ricerche, Piazza L. da Vinci 32, 20133 Milano, Italy

[6] Institut de Physique, Ecole Polytechnique Fédérale de Lausanne, 1015 Lausanne, Switzerland

* These authors contributed equally: Y. Luo, S. Sheng and A. Schirato

+ Authors to whom correspondence should be addressed.

giuseppe.dellavalle@polimi.it and mgarg@fkf.mpg.de

**Probing and manipulating the spatiotemporal dynamics of hot carriers in nanoscale metals is crucial to a plethora of applications ranging from nonlinear nanophotonics to single-molecule photochemistry. The direct investigation of these highly non-equilibrium carriers requires the experimental capability of high energy-resolution (~ meV) broadband femtosecond spectroscopy. When considering the ultimate limits of atomic-scale structures, this capability has remained out of reach until date. Using a two-color femtosecond pump-probe spectroscopy, we present here the real-time tracking of hot carrier dynamics in a well-defined plasmonic picocavity, formed in the tunnel junction of a scanning tunneling microscope (STM). The excitation of hot carriers in the picocavity enables ultrafast all-optical control over the broadband (~ eV) anti-Stokes electronic resonance Raman scattering (ERRS) and the four-wave mixing (FWM) signals generated at the atomic length scale. By mapping the ERRS and FWM signals from a single graphene nanoribbon (GNR), we demonstrate that both signals are more efficiently generated along the edges of the GNR — a manifestation of atomic-scale nonlinear optical microscopy. This demonstration paves the way to the development of novel ultrafast nonlinear picophotonic platforms, affording unique opportunities in a variety of contexts, from the direct investigation of non-equilibrium light-matter interactions in complex quantum materials, to the development of robust strategies for hot carriers harvesting in single molecules and the next generation of active metasurfaces with deep-sub-wavelength meta-atoms.**

Nonradiative decay of optically excited plasmons in nanoscale metals can produce hot carriers – highly energetic electrons and holes whose energy distribution deviates significantly from the equilibrium distribution[1-5]. Efficiently harvesting the energy of these hot carriers is the key to a broad range of emerging applications, e.g., photocatalysis[6,7], photovoltaics[8], and below-bandgap photodetection[9]. Moreover, photogenerated hot carriers can be exploited to modulate nonlinear optical processes in plasmonic[10] or semiconductor[11] nanostructures by all-optical means and at an ultrahigh speed (~ 10 THz), a key functionality for the next generation of nanophotonic devices based on active metamaterials[12].

Albeit the exploitation of hot carriers from nanoscale systems has tremendous potential, optimizing and realizing nanostructures tailored for the above-mentioned applications continues to be challenging. The primary rationale behind this roadblock is the inability to directly investigate hot carriers at their intrinsic spatial length (~ Å), time (~ fs) and energy scales (~ eV)[13,14], due to the difficulty in achieving all the three resolutions simultaneously in state-of-the-art experiments.

Recent advances in combining ultrashort laser pulses with scanning probe microscopy techniques have enabled attaining femtosecond temporal and atomic-scale spatial resolutions, simultaneously[15-23]. However, probing the hot carrier dynamics also necessitates achieving this capability over a broad spectral range[4] (eV) with high energy resolution (meV)[24,25]. The broad spectral range and high energy resolution are crucial as hot carriers exhibit a complex time-dependent behavior, across a wide range of energies, and capturing the subtleties of this spectro-temporal evolution is critical to understanding and manipulating their dynamics.

Here we introduce atomic-scale broadband femtosecond nonlinear spectroscopy and use it to directly probe the hot carrier dynamics at the atomic length and femtosecond time scales in the plasmonic picocavity[26] of a scanning tunneling microscope (STM). Hot carriers generated by the nonradiative decay of photoexcited localized surface plasmons (LSPs) were characterized via the emitted anti-Stokes spectrum over a broad spectral range (~ eV) with ~ 1 meV energy resolution. In order to dynamically control this emission from the picocavity, we performed a two-color pump-probe experiment, where the pump laser pulse controls the hot carrier density, and the spectrally separated probe pulse gives rise to anti-Stokes electronic resonance Raman scattering (ERRS) starting from the out-of-equilibrium carrier distribution induced by the pump pulse. The unique design of our two-color experiment allows the concurrent detection of atomically localized four-wave mixing (FWM) signals, enabling a precise temporal clocking of the hot carrier dynamics.

We performed atomic-scale microscopy of the anti-Stokes and FWM signal intensities in a single graphene nanoribbon (GNR)[27]. The anti-Stokes as well as FWM signals were dramatically enhanced at the edges of the GNR compared to its bulk, which can be attributed to the higher local density of states (DOS) at the edges[27-29]. The enhanced FWM signals at the GNR edges also connote to the fact that higher local DOS leads to a higher nonlinear susceptibility ($\chi^{(3)}$), which varies at the atomic scale[30]. This observation opens new avenues for atomic-scale nonlinear optics, and for the disclosure of a new topic: ultrafast nonlinear picophotonics.

## Nonlinear photonics in a plasmonic picocavity

In our experiments, ultrashort laser pulses (wavelength range, $\lambda$ ~ 715-725 nm, pulse duration, $\tau$ ~ 80 fs) illuminate a plasmonic tunnel junction formed between a Au nanotip and a Au(111) surface of an STM, as schematically shown in Fig. 1a (see Fig. S1 and SI Section I for further details). The interaction with the ultrashort laser pulses generates non-equilibrium hot carriers in the plasmonic tunnel junction, whose energy distribution deviates significantly from the equilibrium Fermi-Dirac distribution. These hot carriers evolve on the femtosecond timescale and are spatially localized at the tunnel junction, as shown by the evaluated photo-absorption pattern (inset of Fig. 1a, see SI Section II-III for further details). The inelastic scattering of the hot carriers with the incident laser pulse leads to the generation of photons whose energies are higher than that of the exciting laser pulses (Fig. 1b), producing anti-Stokes signal.

The anti-Stokes spectrum measured from the picocavity for exciting laser pulses of ~ 37 pJ energy is shown by the blue curve in Fig. 1c. The LSP resonance spectrum measured via electroluminescence by applying a bias of 3 V to the plasmonic picocavity, in the absence of the exciting laser pulse, is shown by the red curve in Fig. 1c. To enable a direct comparison between the two spectra, we intentionally kept the bandwidth of our laser pulses narrower (~ 25 meV) in this measurement, so that the spectrum of the laser pulses did not influence the spectral distribution of the generated hot carriers. We note that the bias voltage applied to the STM junction was kept low (~ 100 mV) when measuring the anti-Stokes spectra, to avoid electroluminescence in the interested spectral range[31-34]. The simulated local photonic density of states (LPDOS) of the plasmonic picocavity, shown by the green curve in Fig. 1c (see SI Section II-IV for the details of the calculations), exhibits a reasonable agreement with the measured spectra of anti-Stokes signal[35] as well as of the LSP resonance[36].

The spatial confinement of the hot carriers in the picocavity was investigated by controllably increasing the size of the cavity. The variation of the spectral intensity of the anti-Stokes signal measured as a function of the increasing tip height ($\Delta z$) is shown in Fig. 1d. The spectral intensity almost vanishes on a relative increase of the cavity size by ~ 4 Å. The variation of the anti-Stokes signal can be fitted with an exponential function, $I_{aS} \propto \exp(-k\cdot\Delta z)$, yielding a decay constant of $k$ ~ 0.78 $Å^{-1}$. This observation clearly points out that the hot carriers responsible for the anti-Stokes signal are confined within the localized plasmonic hotspot between the nanotip and the Au(111) surface.

There are three plausible underlying mechanisms which could lead to anti-Stokes signal from hot carriers: (i) multiphoton absorption ($n \geq 2$) from the exciting laser pulses, followed by interband recombination of the carriers in the conduction and valence bands of Au, giving rise to anti-Stokes emission[37,38]; (ii) intraband recombination of the hot carriers in the Au conduction band, which can be treated as a two-dimensional hot electron gas[39]; and lastly (iii) electronic resonance Raman scattering from the hot carriers[37].

To elucidate the physical mechanism underlying the anti-Stokes signal from the hot carriers, we performed an excitation fluence dependence experiment, where the anti-Stokes spectra were measured as a function of the increasing energy of the exciting laser pulses. The spectrally integrated intensity of the anti-Stokes signal exhibits a quadratic dependence on the laser pulse energy, as shown in Fig. 1e.

The plasmonic response of the cavity as shown in Fig. 1c indicates that the interband transitions (threshold ~ 500 nm)[40,41] between the conduction (*sp*-band) and the valence band (*d*-band) are unlikely, as they would fall in a different spectral range. In the case of intraband recombination, the power-law exponent in the power-scaling experiment should vary linearly with the energy of the photons in the anti-Stokes spectra[39]. However, Fig. S2 in the SI shows negligible variations in the power-law exponent with respect to the measured photon energies, indicating that intraband recombination of the hot carriers is not the mechanism responsible for the anti-Stokes signal. The third mechanism, electronic Raman scattering, is consistent with our observations. In this scenario, the spectral intensity of the anti-Stokes signal is proportional to both the population of hot carriers and the power of the incident laser[37,42]. Since the population of hot carriers is also proportional to the incident laser intensity, eventually, the spectral intensity of anti-Stokes signal exhibits a quadratic dependence on the laser power. The electronic states available to hot electrons and holes in Au form a continuum, thus, the electronic Raman scattering occurs under resonance conditions, making 'electronic resonance Raman scattering' (ERRS) the precise description of the physical mechanism behind the anti-Stokes signal observed in this study. This interpretation is further corroborated by the model we developed to describe the measured signals (comprehensive details can be found in the SI), suggesting that the anti-Stokes signal from the picocavity can indeed be rationalized by the ERRS mechanism. The simulated power-scaling of the anti-Stokes intensity upon single-pulse excitation (green curve in Fig. 1e) precisely follows the experimentally measured quadratic dependence (black dots and red curve in Fig. 1e).

## Broadband femtosecond nonlinear optical spectroscopy at the atomic scale

To time resolve the relaxation dynamics of hot carriers in the picocavity, we performed a broadband two-color pump-probe experiment, as schematically shown in Fig. 2a. Laser pulses with the spectral range of ~ 830-870 nm (~ 30 fs), hereafter, referred to as pump pulse, were used to photoexcite the hot carriers; whereas laser pulses with the spectral range of ~ 715-750 nm (~ 30 fs), hereafter, referred to as probe pulse, were used to track the time evolution of the hot carriers by ERRS (see Methods). The spectral gap between the pump and probe pulses ensures that the pump pulse effectively excite the hot carriers without intervening with the measured anti-Stokes signal, which is exclusively triggered by the probe pulse.

The time evolution of the non-equilibrium distribution of the hot carriers as a function of the delay between the pump and probe pulses is pictorially depicted in Fig. 2a. At negative time delays ($\tau < 0$), when the probe pulse precedes the pump pulse, there is barely any change in the distribution of the hot carriers. In contrast, at zero and positive time delays ($\tau \geq 0$), the probe pulse enables probing the density and the energy distribution of the hot carriers excited by the pump pulse through the time-evolving anti-Stokes spectra. In addition, at zero delay, nonlinear optical processes (three-wave mixing and four-wave mixing (FWM)) can occur between the pump and probe pulses in the picocavity. In the sketch of Fig. 2a, we specifically illustrate the FWM signal (blue arrow) produced by two interactions with the probe pulse (upward transition, green arrows) and one interaction with the pump pulse (downward transition, orange arrow), as this is the only wave-mixing process that can be observed within the spectral range of interest, given the bandwidths of the pump and probe pulses, as shown in Fig. 2b.

The contributions of the FWM (~ 630 – 645 nm spectral region) and the hot carrier signal (~ 660 – 690 nm spectral region) in the anti-Stokes spectra were investigated by individually varying the fluence of the pump and probe pulses at zero time delay between them. When increasing the fluence of the pump pulse, while keeping the fluence of the probe pulse fixed, both the FWM and hot carrier contributions exhibit a linear dependence, as shown in Fig. 2c. Since the pump pulse is responsible for exciting the hot carriers in the picocavity, thus, its linear dependence on the hot carrier contribution is justifiable, as also predicted by our simulation (green curve in Fig. 2c). Similarly, the FWM signal, which involves one interaction with the pump pulse, i.e. stimulating a downward transition ensuing two interactions with the probe pulse, also exhibits a linear dependence on the pump fluence. The FWM signal intensity can be expressed as: $I_{\mathrm{FWM}} \propto \left|\chi^{(3)} E_{probe} E_{probe} E_{pump}^{*}\right|^{2}$, where $E_{probe}$ and $E_{pump}$ are the electric fields of the probe and pump pulses, respectively. $\chi^{(3)}$ is the third-order nonlinear susceptibility.

In contrast, when varying the power of the probe pulses ($P_{Probe}$), the spectral intensity of the FWM contribution ($I_{FWM}$) in the anti-Stokes spectra varies quadratically ( $I_{FWM} \propto P_{\mathrm{Probe}}^{\alpha}$ , α ~ 2.1, blue curve in Fig. 2d), which is consistent with the two interactions with the electric field of the probe pulses in the measured FWM process. However, the spectral intensity of the hot carrier contribution in the anti-Stokes spectra shows a near-linear dependence (α ~ 1.1) on the fluence of the probe pulses (red curve in Fig. 2d), differing from the quadratic dependence observed in the single-pulse power-scaling experiment shown in Fig. 1e.

The fluence dependence measurements further substantiate our interpretation of the electronic resonance Raman scattering as the mechanism behind the anti-Stokes signal. Here, the role of the probe pulse is just to perform ERRS from the hot carriers pre-excited by the pump pulse. In the single-pulse experiment (Fig. 1e), hot carriers are both generated and undergo ERRS by the same pulse, thus, leading to a quadratic power-scaling. However, in the two-color pump-probe experiment, the hot carriers are pre-generated by the pump pulse, resulting in a much less nonlinear dependence (α ~ 1.1) in the power-scaling experiment with the probe pulses, as also confirmed by the simulations, green curve in Fig. 2d.

The spatial extent of the localization of the FWM and hot carrier signals in the anti-Stokes spectra at zero time delay between the pump and probe pulses was explored by controllably increasing the tip-sample distance. Figure 2e shows the variation in the spectrally integrated intensity of both the FWM and hot carrier signals as a function of increasing tip height, achieved by decreasing the tunneling current in the constant current operation mode of the STM. Both the FWM and hot carrier signals decrease dramatically upon increasing the tip height. The FWM signal decays with a constant of ~ 1.2 $Å^{-1}$, which is swifter than the decay rate of the hot carrier signal of ~ 0.8 $Å^{-1}$. The swifter decay of the FWM signal suggests higher sensitivity to the changes in the size of the picocavity in contrast to the hot carrier contribution. This is because FWM depends on the fourth- and second-powers of the locally enhanced probe and pump electric fields, $I_{\mathrm{FWM}} \propto \left|\chi^{(3)} E_{probe} E_{probe} E_{pump}^{*}\right|^{2}$, while the hot carrier contribution (ERRS signal) depends quadratically on the electric fields of both the pump and the probe pulses.

## Energy dependent relaxation dynamics of the hot carriers

A series of anti-Stokes spectra measured as a function of the delay between the pump and probe laser pulses is shown in Fig. 3a. Figure 3b shows anti-Stokes spectra at several representative delays between the pump and probe pulses. The spectral feature at ~ 640 nm observed at the zero delay between the pump and probe pulses, also indicated by red arrow in Fig. 3b, results from the FWM process between the two pulses. This nonlinear FWM signal generated by the pump and the probe pulses facilitates the determination of the absolute time delay between the pulses in the experiment, with the maxima in the position of the FWM signal signifying the absolute time zero in the measurements[43,44]. This enables us to clock the anti-Stokes signal from the probe pulses following the pump-driven excitation of the hot carriers.

A comparison between the temporal cross-cuts of the hot carrier signal at ~ 665 nm and the FWM signal from the spectral region at ~ 640 nm, obtained from the pump-probe measurement, is shown in Fig. 3c. The probe pulses lead to anti-Stokes scattering from the hot carriers generated by the pump pulses almost instantaneously, limited only by the time-resolution of the experiment, which is ~ 30 fs, as indicated by the gray curve in Fig. 3c (see Methods). An exponential fitting of the decay profiles of the measured anti-Stokes spectra (thick red curve in Fig. 3c) reveals the relaxation time of the hot carriers responsible for the anti-Stokes signal at ~ 665 nm to be approximately 149 fs.

Figure 3d shows the simulated time-resolved anti-Stokes spectra, that we modelled by combining a rate-equation-like description of the ultrafast dynamics of hot electrons with a formulation of the ERRS cross-section of the picocavity (see SI section IV-V for the details). For the hot carrier dynamics, we built upon the well-established Three-Temperature Model[45], and extended it to a spatio-temporal model to account for the peculiar ultrafast spatial diffusion of carriers in the picocavity region. For the ERRS, we adapted approaches previously reported[35,46,47] to assess the broadband anti-Stokes Raman scattering. The details of our modelling approach are provided in the SI. Our dynamical simulations exhibit a reasonable agreement with the measured anti-Stokes ERRS signal, both in terms of the spectral shape (besides the FWM fingerprint, not included in our model) and the temporal evolution. In particular, a temporal cross-cut at ~ 665 nm from the simulation is shown in Fig. 3e. Fitting the exponential decay profile reveals a thermalization time of the hot carriers to be ~ 112 fs, which matches reasonably with the measured thermalization time of ~ 149 fs. Our model indicates that the observed signal is dominated by the contribution arising from the nonthermalized hot carriers, while the one from thermalized hot carriers (i.e., closer to the Fermi level) following electron-electron scattering is drastically reduced, due to their spatial diffusion away from the picocavity active region (see SI Section VI for details). A maximum increase in the electronic temperature of mere ~ 30 K is predicted, following a peculiar spatiotemporal dynamics in the picocavity, effectively washing out the contribution of the thermalized hot carriers in the measured anti-Stokes signal. It is worth mentioning that since the temporal cross-cut in the FWM region (~ 630 – 645 nm spectral region) also has an underlying contribution from the hot carriers, hence, it also contains information about their relaxation times, as evident from the relatively broad signal at ~ 640 nm (blue curve in Fig. 3c) compared to the accessible time resolution in the measurements (gray curve, see also Methods). Isolating the FWM contribution in the anti-Stokes spectra is possible by systematically introducing linear positive dispersion in the pump and the probe pulses (Fig. S4 in the SI).

As noticeable from the pump-probe measurement shown in Fig. 3a, the relaxation time of the hot carriers are energy dependent. A spectrally resolved exponential fitting of the temporal cross-cuts of the anti-Stokes spectra reveals the energy dependent relaxation times of the carriers, as shown in Fig. 3f. Hot carriers with an energy of ~ 380 meV (~ 600 nm) thermalize with a relaxation time of ~ 100 fs, while those with an energy of ~ 150 meV (~ 675 nm) exhibit a much longer relaxation time of ~ 200 fs. High-energy hot carriers relax much faster compared to the low-energy hot carriers due to the greater availability electron-electron scattering channels at higher energies[48].

## Atomic-scale microscopy of hot carrier distribution in a single GNR

Lastly, we map the hot carrier and FWM intensity distributions in a single seven-atom-wide graphene nanoribbon (7-AGNR) on the Au(111) surface. We chose GNRs for the study, due to their potential in the development of the next generation of molecular nanoelectronics, where it would be crucial to understand and control hot carrier dynamics at the atomic length scales.

Figure 4a shows the constant current STM topography of a single GNR. The simultaneously measured spatial variation of the spectral intensity of the hot carrier and the FWM signals in the anti-Stokes spectra are shown in Fig. 4b and 4c, respectively. The anti-Stokes spectra were measured at zero delay between the pump and the probe pulses. Both hot carrier and FWM signals are dramatically enhanced at the edges of the GNR compared to its bulk, as also evident from the line profiles shown in Fig. 4d. This enhancement is consistent with the known higher local electronic DOS at the edges of GNRs[29], as also confirmed by differential conductance measurements (see Fig. S3 in the SI). In addition, simulations were performed (see SI Section VIII) to exclude purely photonic (or geometric) effects, i.e. modified LPDOS within the cavity, at the origin of this observation.

A higher DOS would make available a larger pool of carriers which can be photoexcited by the incident pump pulses, leading to a stronger anti-Stokes signal from the hot carriers at the edges of the GNR compared to its bulk (Fig. 4b). In close analogy to the hot carrier signal in the anti-Stokes spectra, the higher DOS at the edges of the GNR could enhance the nonlinear susceptibility to the incident light. As a result, the nonlinear response linked to the third-order nonlinear susceptibility ($\chi^{(3)}$) exhibit atomic-scale variations within the GNR. This study attests to the critical role of the local DOS in facilitating efficient hot carrier generation and in enhancing the nonlinear optical response in single molecules.

Time-resolved dynamics were measured at various places over the GNR. Since the GNR lies flat on the Au(111) surface, it is electronically coupled to it, thus, making the dephasing times of the excited plasmons and eventually the hot carriers very similar to those observed on a clean Au(111) surface[18]. However, for molecules which are electronically decoupled from the metallic surface, a different hot carrier dynamics is expected. This dynamics would uniquely reflect the contributions of the various electronic levels of the molecules[49] and will be the focus of a future work.

## Discussion and Conclusions

We have introduced broadband (~ eV) femtosecond nonlinear optical spectroscopy and microscopy at the atomic length scale, and have used it to demonstrate the concept of ultrafast hot-carrier mediated modulation of anti-Stokes electronic resonance Raman scattering in a plasmonic picocavity. Our setup enables the direct tracking of the photogenerated hot carriers and mapping of their spatial distribution in a single molecular entity. By conducting a two-color pump-probe spectroscopic measurement, we have determined the relaxation times of the hot carriers in the picocavity to be energy dependent, with high-energy hot carriers relaxing faster than their low-energy counterparts. The atomic-scale mapping of FWM and hot carrier distributions opens the door to accessing the nonlinear optical properties and their associated dynamics in individual molecules and complex quantum materials, which were previously accessible only in ensemble measurements at macroscopic length scales.

Atomic-scale broadband nonlinear spectroscopy is the key to realizing the long-sought after goal of seeing chemistry in motion in single molecules[50]. Furthermore, the technique is ideally suited to probe transient photo-induced ferromagnetism and superconductivity[51,52] at atomic scales, and develop a new generation of ultrafast devices based on active metamaterials[12,53] with picosized meta-atoms.

## Methods

### Sample and tip preparation

All the experiments in the current work were performed in a home-built scanning tunneling microscope (STM) operating in ultra-high vacuum (UHV) conditions (~$5\times10^{-11}$ mbar) and at liquid-helium temperature (~11 K). Au(111) surfaces were prepared by repeated cycles of sputtering with 1.0 keV $Ar^+$ ions, followed by thermal annealing at ~430 °C. Au tips prepared by electrochemical etching were used in all the experiments for the plasmonic enhancement. To fabricate the graphene nanoribbons, 0.5 monolayer of 10,10′-dibromo-9,9′-bianthryl (DBBA) molecules were sublimated on the clean Au(111) substrate held at room temperature. The 7-armchair graphene nanoribbons (7-AGNRs) were obtained by post-annealing the sample at 200 °C for 10 minutes, then at 400 °C for another 10 minutes[54].

### Optical Setup

The ultrafast laser system used in the current work is a Ti:Sapphire oscillator (Element™ 2, Newport Spectra-Physics) which produces laser pulses of ~ 6 fs duration with a bandwidth spanning from 650 nm to 1050 nm at a repetition rate of ~ 80 MHz. Laser pulses with reduced spectral range were generated by bandpass filtering the broadband laser pulses from the oscillator, and compressed using chirped dielectric mirror pairs (see Fig. S4 for details). A precise delay stage (N-565, Physik Instrumente) was used to control the delay time between the pump and probe pulses. An achromatic lens (diameter: 50 mm; focusing length: 75 mm) was mounted inside the UHV chamber to focus the laser beams onto the apex of the Au tip. The dispersion accumulated by the laser pulses on passing through several dispersive elements in the setup (achromatic lens and UHV window) was pre-compensated by multiple reflections-off a pair of chirped dielectric mirrors. A second harmonic generation based fringe resolved autocorrelator (FRAC) with an ~ 20 μm thick BBO crystal was used to measure the duration of the pump and probe pulses with identical dispersion as in the optical path to the STM junction. The duration of the pump and the probe pulses were measured to be ~ 30 fs (gray curve in Fig. 3c). The anti-Stokes signal was collected through the same achromatic lens and then focused into the entrance slit of a spectrometer (Kymera 328i, ANDOR) and detected by a thermoelectrically cooled charge coupled device (iDus 416, ANDOR). A schematic of the experimental setup is shown in Fig. S1 in the supplementary materials.

### Modelling ultrafast hot carrier dynamics

The non-equilibrium dynamics of the hot carriers was described numerically by an extended Three-Temperature Model[45] (3TM). In general, the 3TM is a semiclassical rate-equation model which details the ultrafast relaxation of plasmonic hot electrons in terms of three internal energetic variables: (i) the excess energy stored in a 'nonthermal' portion of the carrier population, featuring energies as high as the absorbed photon ones; (ii) an increased electronic temperature, associated with an excited Fermi-Dirac occupancy distribution; and (iii) the metal lattice temperature. Here, given the peculiar electromagnetic mode confinement induced by the picocavity, we adapted the 3TM original formulation to account for the ultrafast spatial diffusion of thermalized hot carriers that additional Finite Element Method (FEM)-based simulations showed to be critical to the electron dynamics. By then numerically integrating our reduced 3TM and conveniently combining the contributions arising from both the probe and the pump pulses, we retrieved the ultrafast evolution of the nonequilibrium carriers' occupancy distribution. Further details on our model can be found in the SI.

### Finite-element electromagnetic simulations

A three-dimensional FEM-based model was developed using commercial software (COMSOL Multiphysics, 6.2) to calculate the optical and electromagnetic properties of the plasmonic picocavity. The system was modelled as a nanoparticle-on-mirror geometry, made of a Au nanosphere (mimicking the STM tip) placed ad a sub-nm distance on top of a flat Au substrate. Both dipolar and plane wave excitations were implemented, enabling us to determine various relevant quantities of the cavity electromagnetic behaviour, including the local photonic density of states, the absorption and scattering cross-sections, the photo-absorption spatial patterns and mode volumes. The details of our model are reported in the SI.

## Data availability

The data that support the findings of this study are available from the corresponding authors on request.

## Code availability

The details needed to reproduce the computations have been provided in the "Methods" section and Supplementary Information file.

## Acknowledgments

We thank Wolfgang Stiepany and Marko Memmler for technical support. A.S., G.D.V. and G.C. acknowledge financial support by the European Union's NextGenerationEU Programme with the I-PHOQS

Infrastructure [IR0000016, ID D2B8D520, CUP B53C22001750006] "Integrated infrastructure initiative in Photonic and Quantum Sciences", and from the METAFAST project that received funding from the European Union Horizon 2020 Research and Innovation program under Grant Agreement No. 899673. This work reflects only the author's view, and the European Commission is not responsible for any use that may be made of the information it contains. G.D.V. acknowledges the support from the HOTMETA project under the PRIN 2022 MUR program funded by the European Union – Next Generation EU - "PNRR - M4C2, investimento 1.1 - "Fondo PRIN 2022" - HOT-carrier METasurfaces for Advanced photonics (HOTMETA), contract no. 2022LENW33 - CUP: D53D2300229 0006". A.S. and G.D.V. acknowledge the European Union's Horizon Europe research and innovation programme under the Marie Skłodowska-Curie Action PATHWAYS HORIZON-MSCA-2023-PF-GF grant agreement No. 101153856. A.M.J acknowledges funding from HORIZON-MSCA-2022-PF-01-01 under the Marie Skłodowska-Curie grant agreement No. 101108851.

## Contributions

Y.L., S.S., A.M.J., K.K., and M.G. built the experimental setup, performed the experiments and analyzed the experimental data. A.S., G.C., and G.D.V. designed and performed the theoretical calculations and analyzed the theoretical data. M.G. conceived the project and designed the experiments. All authors interpreted the results and contributed to the preparation of the manuscript.

## Competing interests

The authors declare no competing interests.

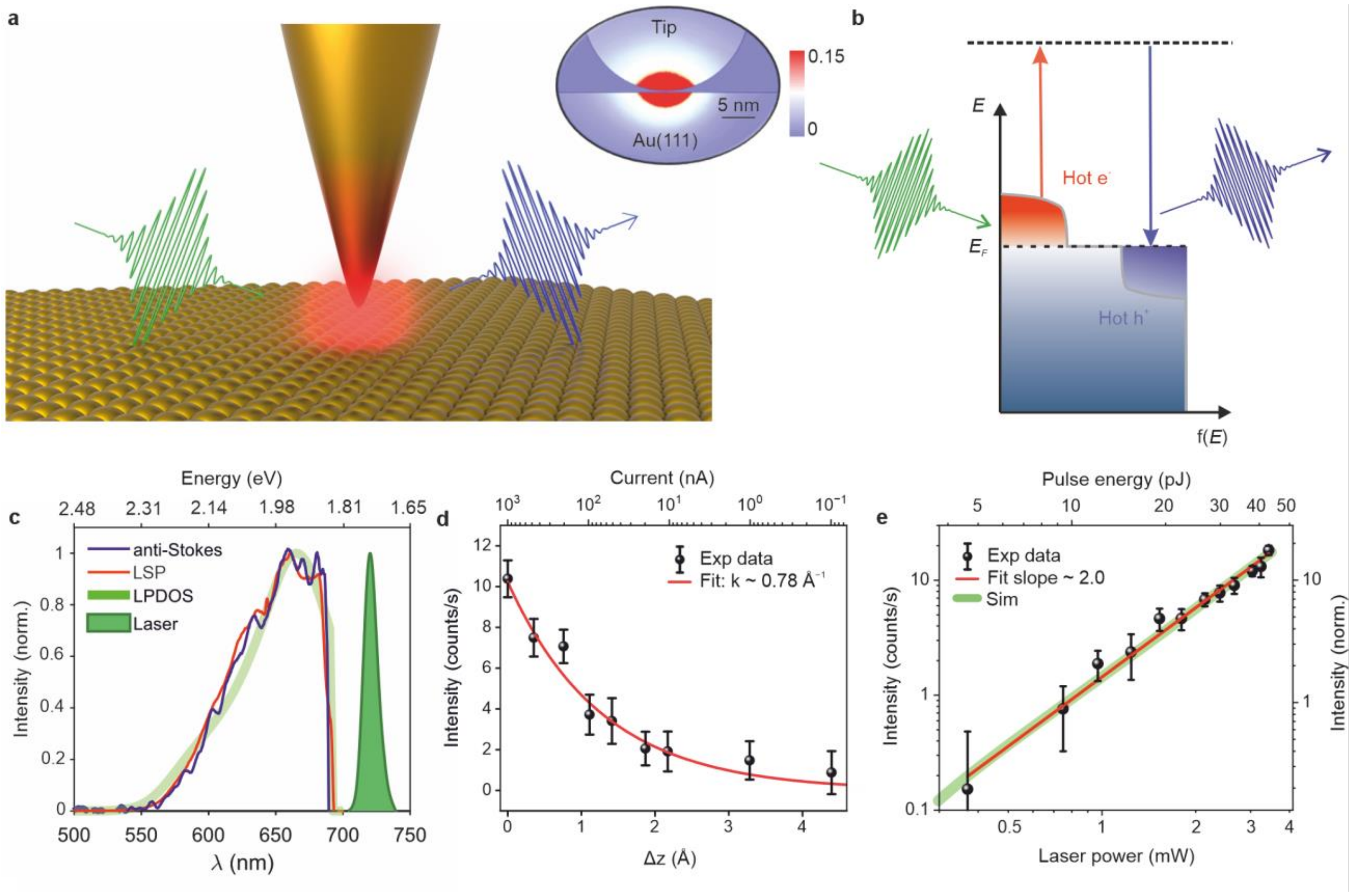

**Fig. 1 | Ultrafast hot-carrier driving of nonlinear optical processes in a plasmonic picocavity. a,** Schematic depiction of the hot carrier photogeneration in the plasmonic tunnel junction of a scanning tunneling microscope (STM) by ultrashort laser pulses (green), ensued by an anti-Stokes electronic Raman scattering (blue). Inset: Simulated spatial distribution of the electromagnetic dissipated power across the picocavity (normalized), showing the extreme localization of the optical modes enabled by the plasmonic picocavity. **b,** Schematic illustration of hot carriers generation and anti-Stokes light emission: energy distribution of the non-equilibrium hot carriers (electrons and holes) under ultrashort laser pulse excitation, leading to anti-Stokes light emission. $E_F$: Fermi energy level. **c,** Comparison of the anti-Stokes spectrum (blue curve) from the plasmonic junction with the local surface plasmonic resonance (red curve), and the calculated local photonic density of states (LPDOS) of the plasmonic picocavity (light green curve). The spectrum of the exciting laser pulses is shown by the filled green curve. Top x-axis represents the spectral axis in eV. **d,** Variation of the measured anti-Stokes signal intensity at ~ 680 nm (black dots) as a function of the increasing plasmonic picocavity size (Au nanotip – Au(111) sample distance) and with the corresponding decreasing tunneling current (top x-axis). An exponential fit of the measured anti-Stokes signal (red curve) yields a decay constant of ~ 0.78 $Å^{-1}$. $\Delta z = 0$ Å in the plot represents the height of the nanotip from Au(111) surface at the tunneling condition of 1 μA at 100 mV. **e,** Variation of the measured anti-Stokes signal intensity (black dots) and simulated LPDOS (light green curve) as a function of

increasing incident laser power, plotted in a dual-logarithmic scale, respectively, and the corresponding quadratic fit of the measured data (red curve). STM junction was operated under constant current mode with the set tunneling current of 500 nA at 100 mV. Parameters of the exciting ultrashort laser pulse: wavelength range: $\lambda \sim$ 715–725 nm, pulse duration ~ 80 fs, power ~ 3 mW. Error bars in **d** and **e** represent the standard deviation from the integrated spectral area.

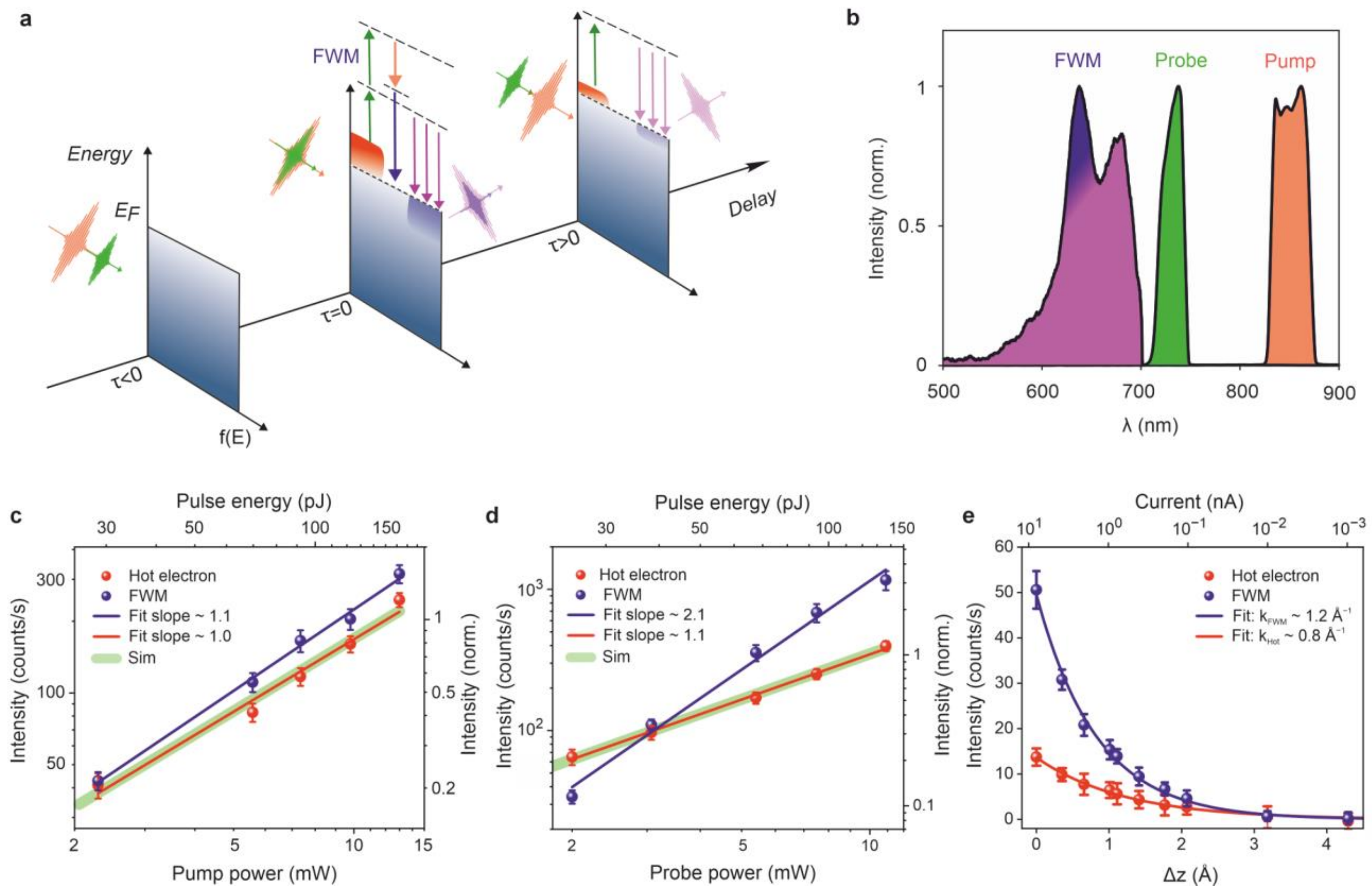

**Fig. 2 | Two-color broadband nonlinear spectroscopy of hot carriers. a**, Sketch of the temporal evolution of the energy distribution of the hot carriers in the picocavity as a function of the pump-probe delay. Left-panel: electron distribution when the probe pulse (green curve) precedes ($\tau < 0$) the pump pulse (orange curve). Middle-panel: hot carrier generation on excitation by the pump pulse followed by anti-Stokes emission (vertical purple curves) upon interaction of the hot carriers with the probe pulse (vertical green-curve, $\tau$=0). Four-wave mixing (FWM) process (vertical blue curve) occurs when the pump (orange) and probe pulses (green) temporally overlap. Right-panel: relaxation of the hot carriers after the pump pulse excitation ($\tau > 0$), which is tracked in real-time by the evolving anti-Stokes signal generated by the probe pulses. **b,** Spectra of the pump pulse (orange curve), probe pulse (green curve), and the anti-Stokes spectrum measured at the temporal overlap (zero delay) between the pump and the probe pulses. The purple and the blue shaded regions in the anti-Stokes spectrum depict the contributions of the hot carriers and the FWM signals, respectively. **c, d,** Variation in the intensity of the hot carrier (red dots) and FWM (blue dots) contributions in the anti-Stokes spectra as a function of the increasing fluence of the pump (**c**) and probe (**d**) pulses, plotted in a dual-logarithmic scale, respectively. Red (blue) curves represent the power law fittings, with the fitted slopes (exponents) mentioned in the figure legends. Green curves show the numerically calculated variation in the intensity of the hot carrier contribution (at 680 nm) in the simulated

anti-Stokes spectra upon change of the fluence of the pulses. Probe and pump laser power were fixed at 3.1 mW and 7.35 mW in **c** and **d**, respectively. **e**, Variation in the intensity of the hot carrier (red dots) and FWM (blue dots) contributions in the anti-Stokes spectra as a function of increasing tip height ($\Delta z$), with the corresponding decreasing tunneling current (top x-axis). Red and blue curves show the exponential fits of the hot carrier and the FWM signals, respectively. $\Delta z = 0$ Å in the plot represents the height of the nanotip from Au(111) surface at the tunneling condition of 8 nA at 100 mV. Probe and pump laser powers were fixed at 3.1 mW and 7.35 mW, respectively. The delay between pump and probe pulses was set to be zero fs in **c**, **d**, and **e**. Error bars in **c**, **d**, and **e** represent the standard deviation from the integrated spectral region.

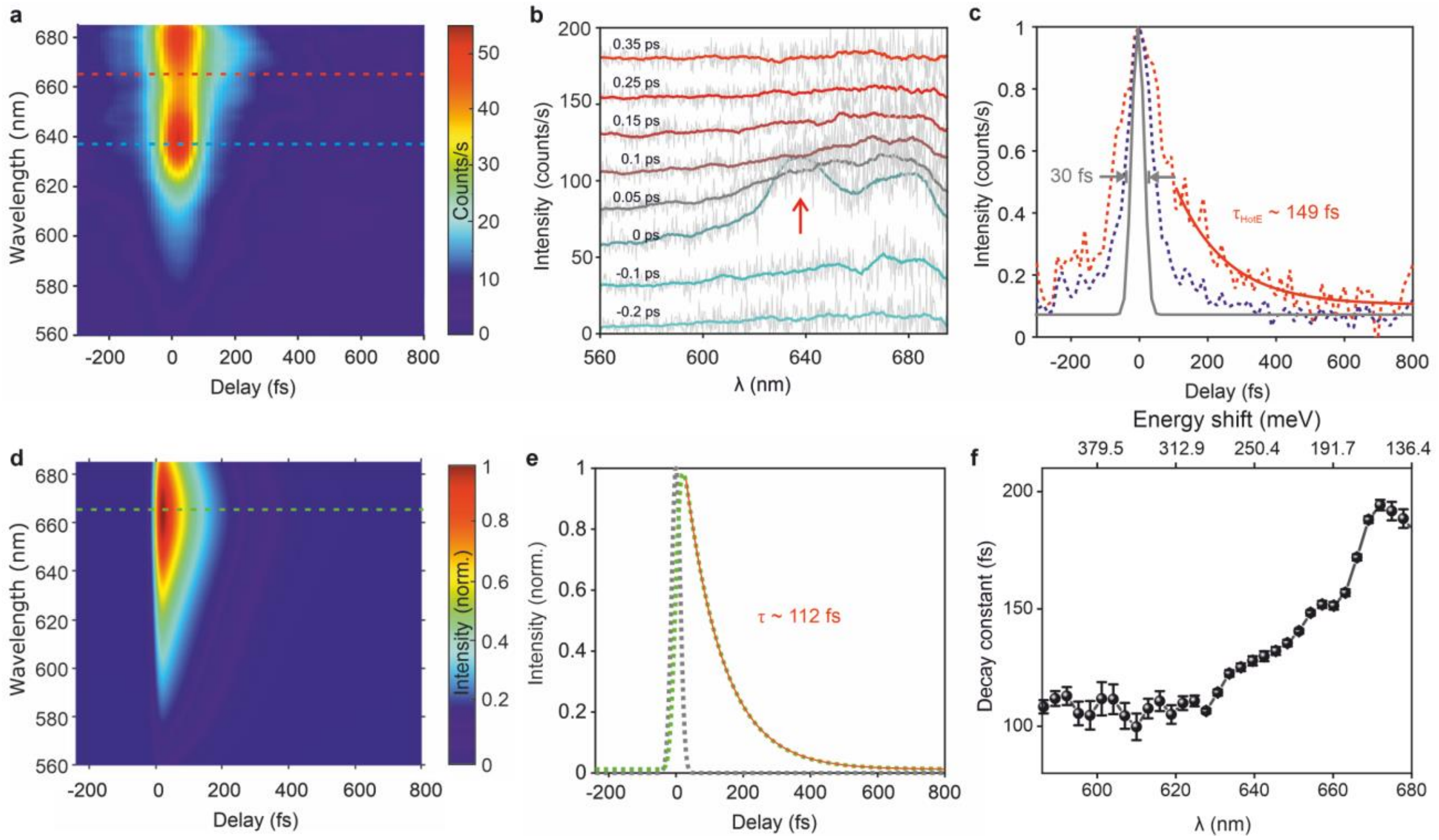

**Fig. 3 | Atomic-scale tracking of hot carrier dynamics. a**, A series of anti-Stokes spectra measured as a function of the delay between the pump and the probe pulses. Pump pulse parameters: λ ~ 830-870 nm, pulse duration ~30 fs, power ~ 5.8 mW (72 pJ). Probe pulse parameters: λ ~ 715-750 nm, pulse duration ~ 30 fs, power ~ 2.6 mW (32 pJ). STM was operated in the constant current mode with a tunneling current of 8 nA and a bias of 100 mV. **b**, Representative anti-Stokes spectra from the measurement shown in **a** at different delays between pump and probe pulses, as annotated on top of each spectrum. **c**, Spectral intensity variation of the anti-Stokes spectra at ~ 665 nm (hot carrier, dashed red curve) and at ~ 640 nm (FWM, dashed blue curve) as a function of the delay between pump and probe pulses, indicated by horizontal dashed red and blue lines in **a**, respectively. The gray curve represents the duration of the probe pulse (see Methods). An exponential fit (solid red curve) of the temporal cross-cut reveals a relaxation time of ~ 149 fs. **d**, Simulated anti-Stokes spectra as a function of the delay between the pump and the probe pulses. The parameters of the pump and probe pulses used in the simulations are identical to the experimental ones. **e,** Temporal cross-cut at 665 nm from the simulation shown in **d** (horizontal dashed green curve)**.** A relaxation time of ~ 112 fs is estimated from the exponential fit of the temporal cross-cut (solid red curve). **f,** Variation of the measured relaxation times of the anti-Stokes signal as a function of various wavelength. Top x-axis represents the energy shift of the anti-Stokes signal with respect to the central wavelength (~ 735 nm) of the probe pulse. The error bars indicate the standard deviation from the fit.

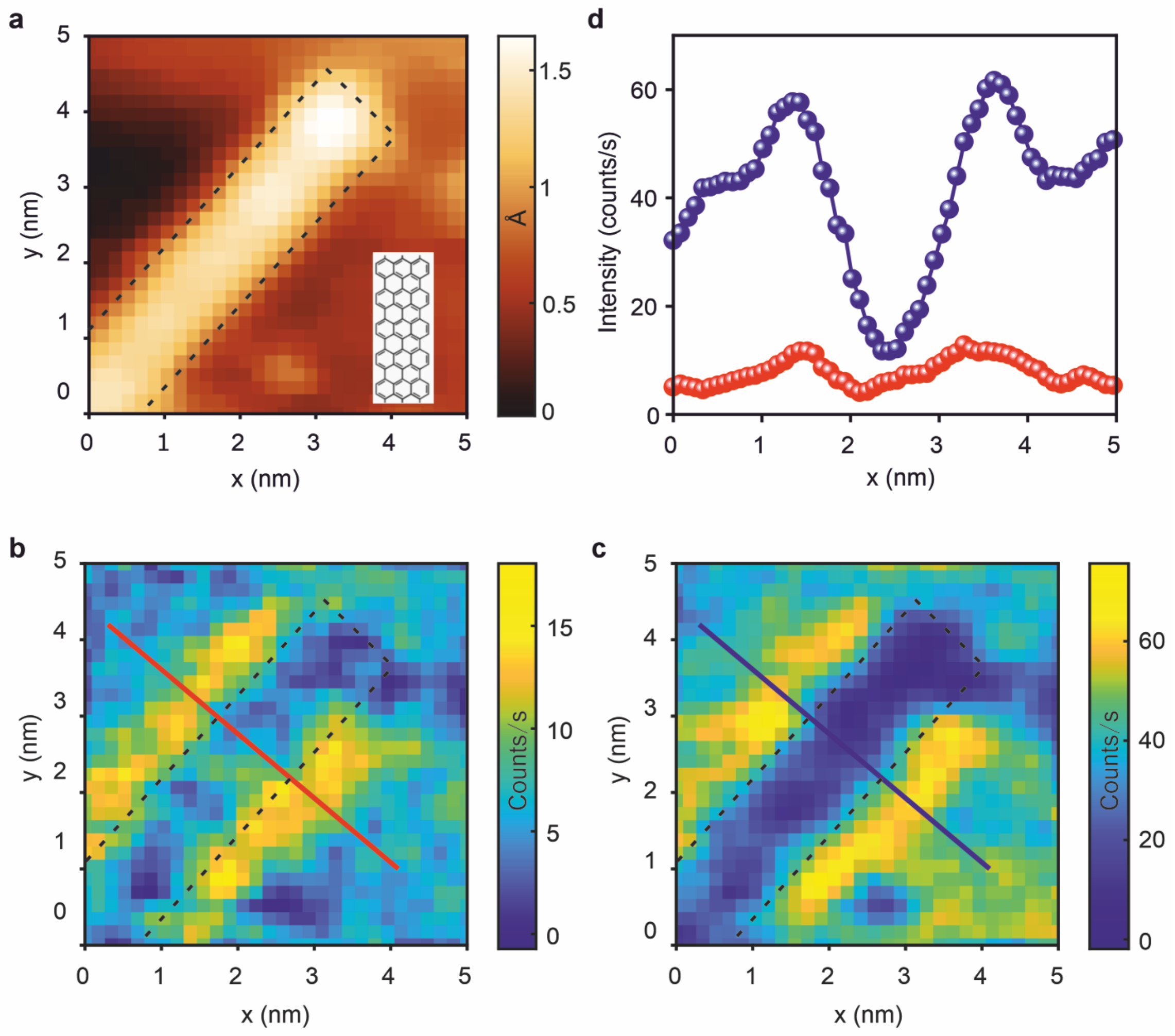

**Fig. 4 | Atomic-scale microscopy of hot carriers and nonlinear susceptibility ($\chi^{(3)}$). a,** STM image of a single seven-atom-wide graphene nanoribbon (7-AGNR) on the Au(111) surface. The color bar represents the relative change in the vertical position of the Au nanotip (in Å) while scanning the GNR in the constant current mode of the STM with a tunneling current of 1 nA at 1 V. Bottom-right inset in **a** shows the chemical structure of the GNR. **b, c,** Spatial variation of the intensity of the hot carrier (**b**) and the FWM (**c**) contributions in the anti-Stokes spectra recorded simultaneously with the topography shown in **a**. The delay between the pump and the probe pulses was set to be zero. The hot carrier and the FWM contributions in the anti-Stokes spectra were evaluated by spectral integration in wavelength range of 665-675 nm and 635-645 nm, respectively. Dashed black rectangles in **a**, **b** and **c** indicate the location of the GNR. **d,** Variations of the hot carrier (red) and FWM (blue) signals intensities along the annotated lines in **b** and **c**, respectively. Probe and pump laser power were fixed at 2.5 mW and 5.5 mW, respectively.

Supplementary Information for

# Visualizing hot carrier dynamics by nonlinear optical microscopy at the atomic length scale

Yang Luo[1,*], Shaoxiang Sheng[1,*], Andrea Schirato[2,3,*], Alberto Martin-Jimenez[1,4], Giuseppe Della Valle[2,5 +], Giulio Cerullo[2,5], Klaus Kern[1,6], Manish Garg[1,+]

[1] Max Planck Institute for Solid State Research, Heisenbergstr. 1, 70569 Stuttgart, Germany

[2] Dipartimento di Fisica, Politecnico di Milano, Piazza L. da Vinci 32, 20133 Milano, Italy

[3] Department of Physics and Astronomy, Rice University, 6100 Main St Houston, Texas 77005, United States

[4] Instituto Madrileño de Estudios Avanzados en Nanociencia (IMDEA Nanociencia), Calle Faraday 9, 28049 Madrid, Spain

[5] Istituto di Fotonica e Nanotecnologie – Consiglio Nazionale delle Ricerche, Piazza L. da Vinci 32, 20133 Milano, Italy

[6] Institut de Physique, Ecole Polytechnique Fédérale de Lausanne, 1015 Lausanne, Switzerland

* These authors contributed equally: Y. Luo, S. Sheng and A. Schirato

+ Authors to whom correspondence should be addressed.

giuseppe.dellavalle@polimi.it and mgarg@fkf.mpg.de

## I. Experimental Details

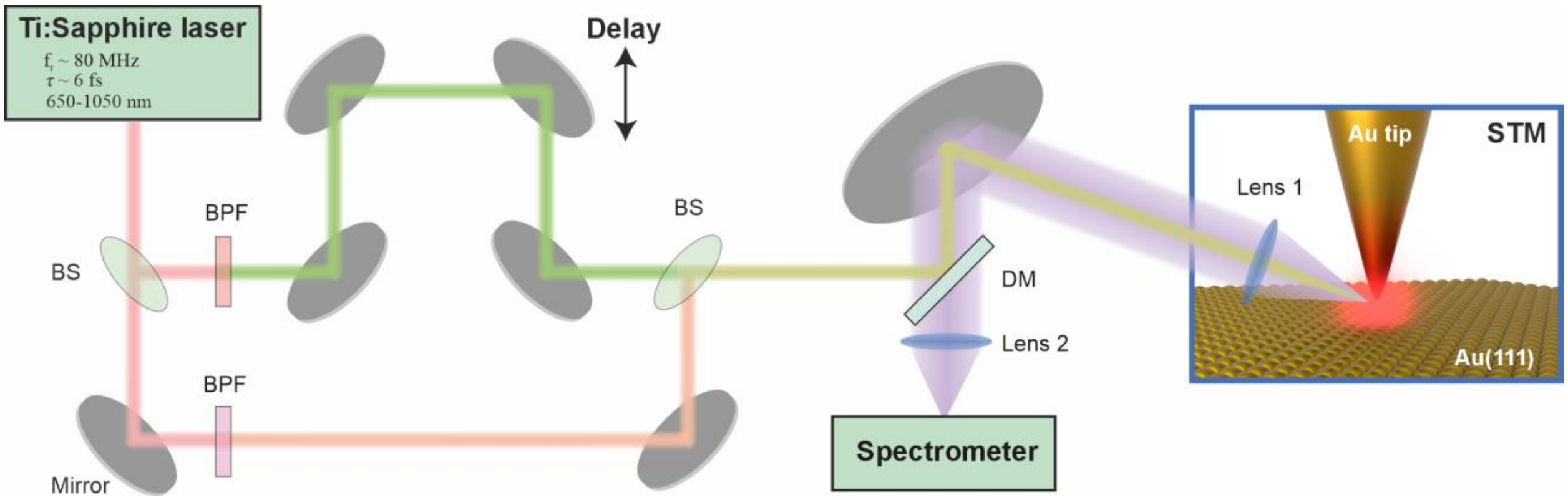

**Fig. S1 | Experimental setup.** In the two-color pump-probe experiment, pump pulse with the spectral range of ~ 830 – 870 nm (pulse duration ~ 30 fs) and probe pulse with the spectral range of ~ 715 – 750 nm (pulse duration ~ 30 fs) were generated by using different bandpass filters in the two arms of the pump-probe setup. In the single-pulse experiments (Fig. 1, main-text), ultrashort laser pulses with the spectral range of ~715 – 725 nm (pulse duration ~ 80 fs) were generated in one arm, while the other arm of the setup was blocked. BS: beam splitter; BPF: bandpass filter; DM: dichroic mirror.

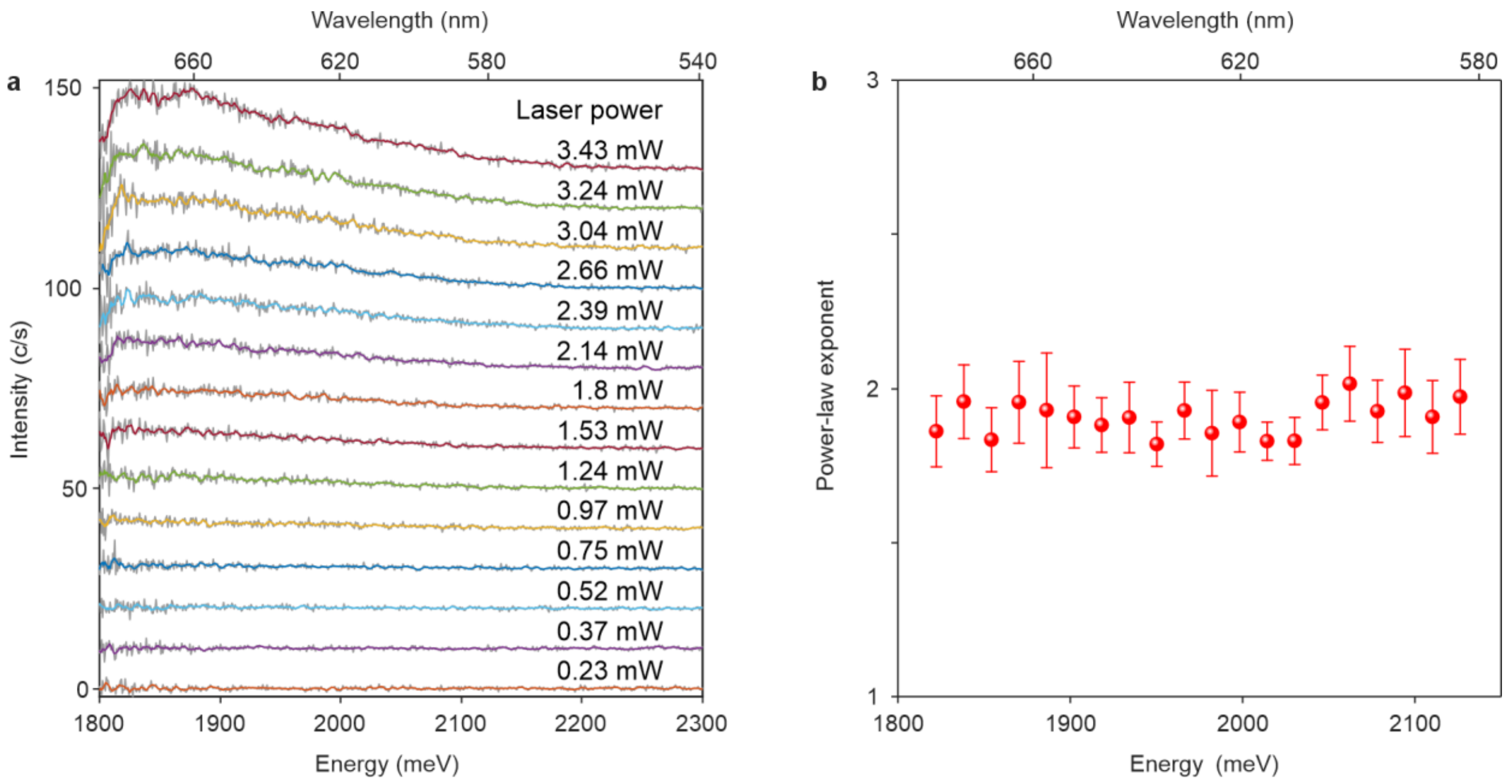

**Fig. S2 | Spectrally-resolved power-law exponent in the anti-Stokes spectra. a**, A series of anti-Stokes spectra measured at various powers of the incident laser pulses, as annotated on top of each spectrum. Ultrashort laser pulses with the spectral range of ~ 715-725 nm (~ 80 fs) were used to generate the anti-Stokes signal, as indicated in Fig. 1 of the main-text. **b**, Energy dependence of the power-law exponent. The spectral intensity at each photon energy (red dots) was obtained by integrating over an energy width of ~16 meV. The error bars indicate the standard deviation from the fit. Top x-axis in **a** and **b** represent the spectral axis in nm.

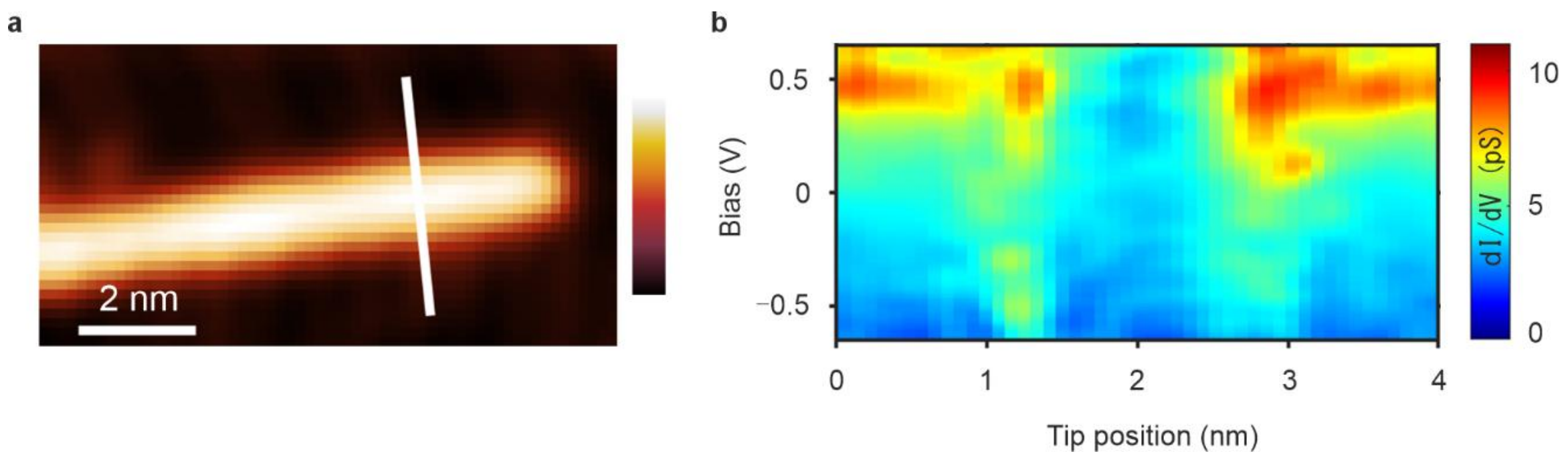

**Fig. S3 | Differential conductance measurements on a single GNR. a**, STM topography of a single GNR on the Au(111) surface, measured in the constant current mode of the STM junction, with a tunneling current of 33 pA and a bias of 1 V. The color scale represents the relative variation in the tip height during the measurement. **b**, A series of differential conductance (dI/dV) spectra measured along the annotated white line in **a** (from top to bottom). The modulation voltage for the measurement was 40 mV at the frequency of 887 Hz. The height of the nanotip during the measurement (open-feedback) was stabilized at a tunneling current of 100 pA and a bias of 1 V. The color scale represents the conductance. The x-axis represents the tip position, whereas the y-axis denotes the biases used for the differential conductance measurement

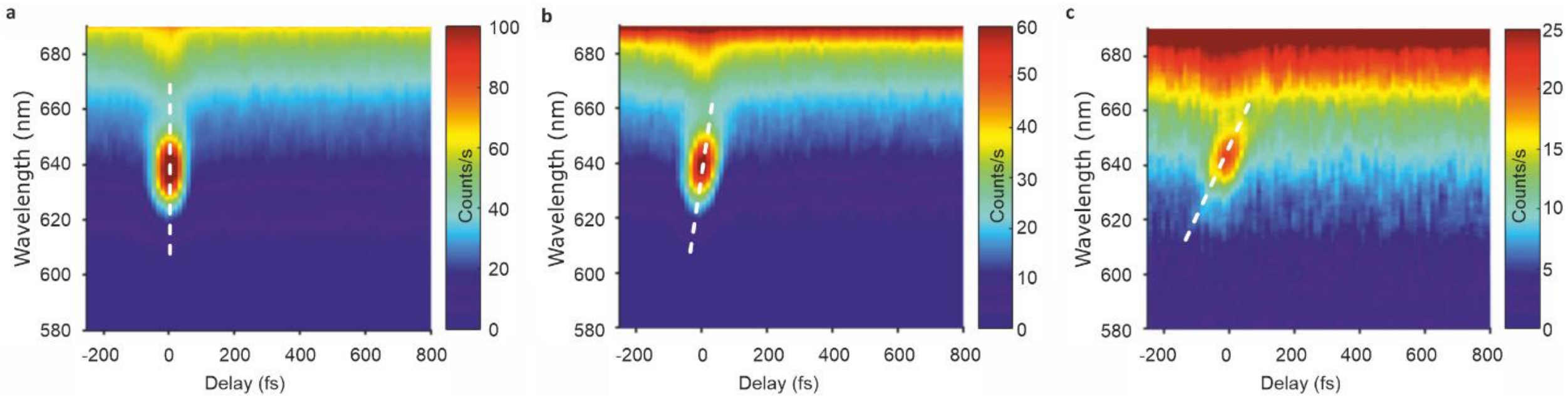

**Fig. S4 | Pump-probe measurements in the plasmonic picocavity as a function of the increasing positive chirp of the pump and probe laser pulses**. **a**, Chirp-free condition. **b**, Positive chirp by adding 16 mm glass in the pump and probe beam paths. **c**, Positive chirp by adding 32 mm glass in both beam paths. STM junction set point was 8 nA at 100 mV for all the measurements. Vertical white dashed-lines in **a**, **b** and **c** show the increasing positive tilt (chirp) in the FWM signal. The two-color pump-probe experiments in the main-text were conducted at the chirp-free condition (**a**).

## II. Calculation of the local photonic density of states (LPDOS) of the picocavity

To calculate the local photonic density of states (LPDOS) of the plasmonic picocavity under investigation, numerical simulations have been performed using the Finite Element Method (FEM)-based software COMSOL Multiphysics (version 6.2). The system was modelled as a three-dimensional (3D) nanoparticle-on-mirror (NPoM) geometry[1], consisting of a spherical Au nanoparticle (mimicking the STM tip) embedded in a homogeneous environment (here air), and placed at a distance $d_{gap}$ above a semi-infinite flat Au substrate. A schematic of the geometry used in the simulations is shown in Fig. S5, where $d_{gap}$ = 0.5 nm, akin to the experimental setup. The NP radius was set to 25 nm, to best match the spectral features of the measured LPDOS. However, minor changes of the simulated spectra were obtained by varying the value of the sphere radius, which only slightly modifies the effective longitudinal size of the cavity.

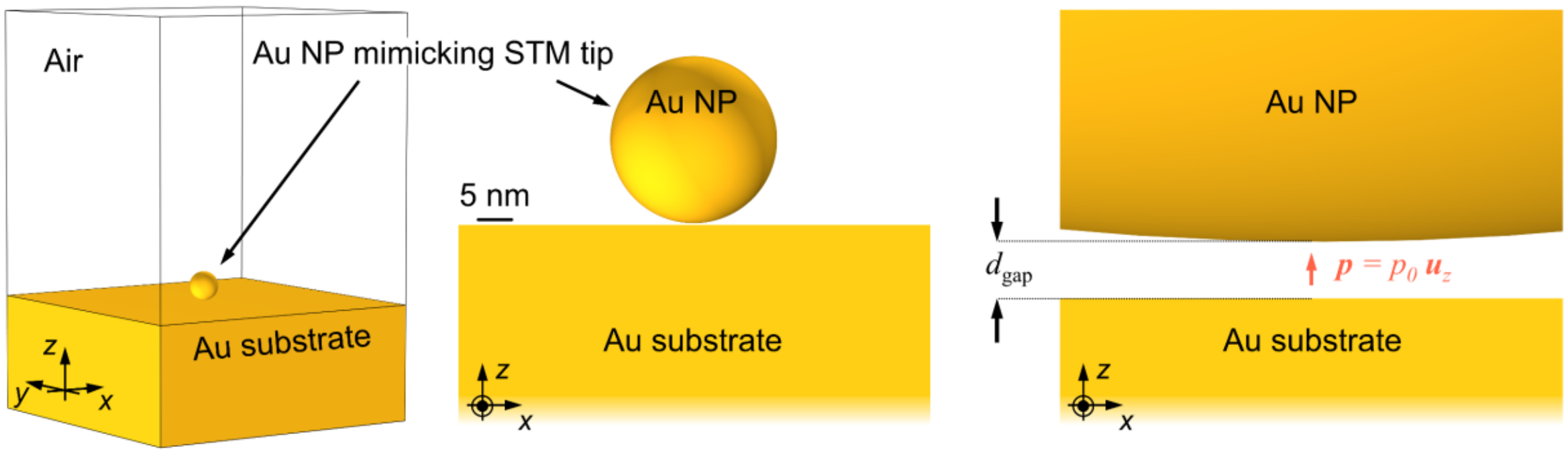

**Fig. S5 | Simulations of the plasmonic picocavity LPDOS**. Schematic of the NPoM geometry used in the simulations. The zoomed-in sketch of the picocavity (rightmost side) shows the point-like dipolar source (in red, vertically oriented) used to calculate the local photonic density of states.

The LPDOS of the system was then numerically derived from its electromagnetic response upon dipolar excitation, following standard approaches. More precisely, a point-like dipole source with unitary dipole moment was introduced inside the sub-nm gap between the NP and the flat metal surface, and aligned to the vertical ($z$-) direction, namely expressed as $\boldsymbol{p} = p_0 \cdot \boldsymbol{u_z}$, with $p_0$ = 1 C·m. In the main simulations, the source was placed at the centre of the cavity, i.e., at $d_{gap}/2$ = 0.25 nm above the metallic surface, and at the same in-plane position as the spherical NP pole. In agreement with previous studies[2], the position of the dipole within the gap in such NPoM configuration only lightly affected the results of the simulations. For the optical properties of Au, the permittivity was described by an analytical expression fitted on experimental data[3], as given in Ref. 4. In addition, to reproduce the broad spectral features in the system optical response, the value of the Drude damping factor $\Gamma_D$ was increased with respect to the nominal coefficient for bulk Au[5], $\Gamma_0$. A good agreement was found for $\Gamma_D/\Gamma_0 = 1.5$. Maxwell's equations were solved in the full-field formalism, by including perfectly matched layers (PMLs) surrounding the physical domain, and scattering boundary conditions (BCs) beyond the PMLs. To ensure numerical accuracy, the tip region was meshed with particularly fine elements, given a minimum (maximum) size of 0.1 (5) nm. Convergence tests indicated the numerical solution to be stable upon further refinement of the meshing.

Once the electromagnetic field distribution induced by a point-like dipole excitation has been calculated across the whole system, the LPDOS $\rho_{phot}(\mathbf{r}, \omega)$, as a function of the position **r** and the dipole source angular frequency ω, can be retrieved according to the following expression[6, 7]:

$$\rho_{phot}(\mathbf{r}, \omega)= \frac{4\varepsilon_0}{2\pi\,\omega\, p_0}\mathrm{Re}[\mathbf{E}(\mathbf{r}, \omega)\cdot \boldsymbol{u}_z], \quad \text{(S1)}$$

(in $s/m^3$) where **r** represents the position of the point-like dipolar source.

Note that the expression above provides the total LPDOS, i.e. the sum of its radiative and non-radiative (namely, absorptive) contributions[8]. To distinguish between these two terms, an alternative approach can be employed to compute $\rho_{phot}$ in terms of the system's scattered (i.e., re-radiated) and absorbed powers, $P_{rad}$ and $P_{abs}$, respectively. Specifically, from the solution of the electromagnetic problem, one can retrieve the following quantities:

$$P_{rad}(\mathbf{r}, \omega) = \int_{\Sigma} \boldsymbol{S} \cdot \mathrm{d}S \quad \text{(S2)}$$

$$P_{abs}(\mathbf{r}, \omega) = \int_{V} Q_{diss}\, \mathrm{d}V \quad \text{(S3)}$$

where $\boldsymbol{S}$ is the Poynting vector, Σ is a surface surrounding the whole system, $Q_{diss}$ represents the absorption Ohmic losses produced within the total volume of metal $V$ (including the spherical NP and the whole Au substrate in our simulations). The two quantities are also considered as functions of the position **r** of the dipolar emitter. The corresponding total extincted power $P_{tot}$ is then simply the sum of these two terms, that can be used to express the LPDOS as follows[6]:

$$\rho_{phot}(\mathbf{r}, \omega)=\left[\frac{P_{tot}(\mathbf{r}, \omega)}{P_{vac}(\omega)}\right]\rho_{vac}(\omega) \quad \text{(S4)}$$

where $\rho_{vac}(\omega) = \omega^2/3\pi^2 c^3$ is the LPDOS in vacuum, and $P_{vac}$ is the power radiated by a point-like dipole source in vacuum given by Larmor's formula[7]. From the relation given in Eq. S4, it becomes straightforward to identify the two, scattering and absorption, contributions in the total LPDOS, by replacing $P_{tot}$ by either $P_{rad}$ or $P_{abs}$, respectively. Such distinction is especially relevant since, in agreement with previous reports[9], the meaningful quantity for our analysis is the solely radiative component of the LPDOS, namely $\rho_{vac}[P_{rad} / P_{vac}]$, as this is the one contributing to the photons detected in our experiments.

As such, in all of the expressions and discussions below, the radiative term of the LPDOS will be used. Hereafter, the quantity $\rho_{phot}$ will therefore refer to the radiative LPDOS of the plasmonic picocavity under investigation.

Finally, note that the expressions given above for the LPDOS depend in general on the position of the point-like dipole source defining the excitation. However, in agreement with previous studies treating equivalent NPoM configurations[2], the results of our simulations showed a minor dependence on the position of the dipolar source. For seek of simplicity, we therefore employed directly the LPDOS obtained for a dipole placed at the precise centre of the cavity in the rest of our models.

### III. Calculation of the far-field optical response of the picocavity

To analyse the interactions between the picocavity and laser light, we modelled the far-field optical behaviour and electromagnetic response of the system upon plane-wave excitation. This allowed us to estimate quantities such as the absorption cross-section and the spatial distribution of the electromagnetic dissipation, relevant to model the photoinduced non-equilibrium carrier dynamics (see Sections below). The simulations were performed on the same 3D geometry discussed above (Fig. S5), and considering the same optical permittivities of the material. Unlike the calculations detailed in Section II, Maxwell's equations were solved in the scattering formalism, with a background field given by the solution of the electromagnetic problem for a planewave excitation of the structure in the absence of the Au NP (i.e., a semi-infinite metal film with a semi-infinite air layer on top). Such background field was evaluated beforehand numerically, using the COMSOL built-in periodic ports to set the electromagnetic excitation, based on the experimental illumination conditions (sketched in Fig. S6a). In particular, monochromatic light with a grazing (defined by the wave-vector from the metal film surface) incident angle of 12° was considered, with TM polarisation (in-plane electric field).

Standard formulas[10] were applied to characterise the global optical response of the picocavity. Specifically, we evaluated the absorption, $\sigma_{abs}$, and scattering, $\sigma_{sca}$, cross-sections of the system, whose simulated spectra are shown in Fig. S6b. A resonant profile is obtained for both cross-sections, with a peak in scattering at around 680 nm (in agreement with the resonant feature observed in the simulated radiative component of the LPDOS) and a slightly blue-shifted one (at ~660 nm) in absorption, which largely dominates over the former (note the distinct vertical axes).

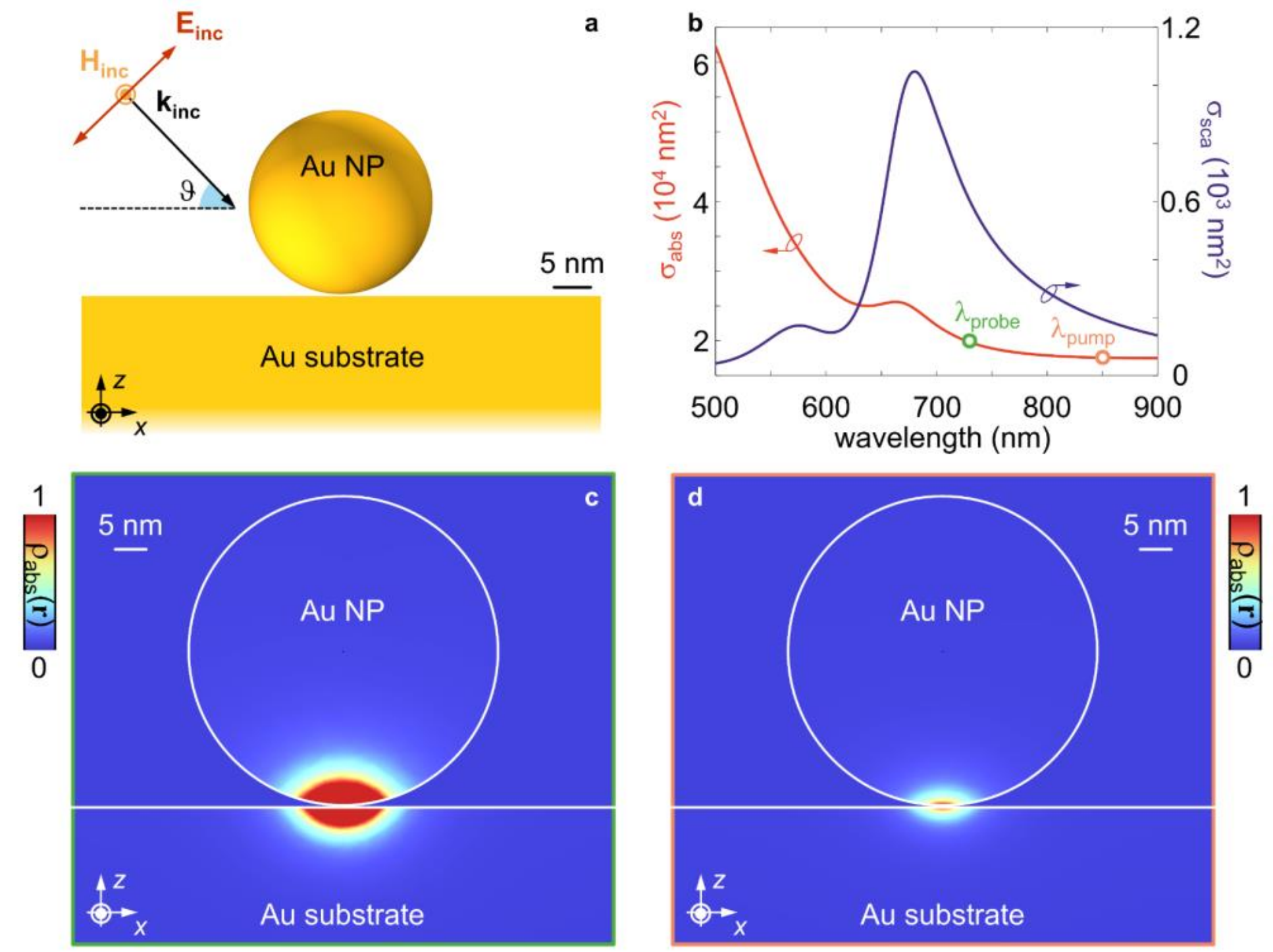

**Fig. S6 | Simulations of the plasmonic picocavity optical response. a,** Sketch of the numerical geometry and planewave illumination conditions considered in the model. **b,** Simulated spectra of the absorption (red, left vertical axis) and scattering (blue, right vertical axis) cross-sections of the picocavity. **c,** Normalized spatial distribution (cross-sectional 2D cut) of the electromagnetic dissipated power $\rho_{abs}(\mathbf{r})$, evaluated at the probe pulse central wavelength (730 nm). **d,** Same as (c) for the pump pulses central wavelength (850 nm). Both maps in (c) and (d) are normalized to the maximum of the dissipated power at the pump wavelength.

Moreover, the quantitative estimation of the full spectrum of $\sigma_{abs}$ allowed us to assess consistently the power absorbed by the NPoM system when excited with either pump or probe pulses (their spectral position being highlighted by colored dots in Fig. S6b). Simulations predict a relatively small (~10%) difference in the $\sigma_{abs}$ at the central wavelengths of the two pulses (850 nm and 730 nm for pump and probe, respectively), sitting relatively far from the region where the metal losses increase substantially, at shorter wavelengths (< 550 nm). In addition, with the 3D numerical model we could inspect the spatial distribution of the induced fields at the two different wavelengths, providing insight into the volume of the optical modes mediating the interaction. In particular, Figs. S6c and S6d show cross-sectional 2D cuts (in the central *z*-*x* plane, at $y = 0$) of the simulated electromagnetic power density $\rho_{abs}(\mathbf{r}, \lambda)$ (defined as the Ohmic losses inside the metal, and normalised to the maximum of the dissipated power at the pump wavelength) across the system, evaluated at the probe (Fig. S6c, 730 nm) and pump (Fig, S6d, 850 nm) pulses central wavelengths, respectively. The numerical results show the extremely localized character of the optical modes excited at the two wavelengths. The intensity of $\rho_{abs}(\mathbf{r}, \lambda)$ rapidly decays when moving away from the vicinity of the sub-nm cavity gap. Based on these spatial maps, we estimated a typical volume $V_{EM}$ of ellipsoidal shape with semiaxes ~11 nm x 11 nm x 4 nm to approximately describe the spatial extent of the optical modes excited at the frequencies of these pulses.

## IV. Modelling anti-Stokes (aS) emission

The power of the generated anti-Stokes (aS) light was calculated following an approach similar to the one proposed in Refs. 11-13. In essence, given the intensity $I$ of an incoming probe beam, the power of the aS light generated from our picocavity as a function of the frequency ω was expressed as $P_{aS}(\omega)=\sigma_{aS}(\omega)I$, where $\sigma_{aS}(\omega)$ represents the resonant electronic aS Raman scattering cross-section of the picocavity. As for any scattering process, $\sigma_{aS}$ scales with the square of the (active, i.e. modal) volume $V_{EM}$ of the probe (that we estimated via the electromagnetic simulations presented in Section III), and it is proportional to the local photonic density of states of the picocavity (LPDOS, refer to Section II for the details on how we computed this quantity). Note that in our calculations of the cavity LPDOS, the minor impact of the position of the emitting dipole on the simulated spectra allowed us to treat a space-independent LPDOS, computed for a dipole source located at the centre of the cavity, and here referred to as $\rho_{phot}(\omega)$. Finally, considering the most general situation in our experiments, where a further laser pulse (the pump), was present, $\sigma_{aS}$ depends also on the pump-probe time delay $t$. Upon these assumptions, we formulated the following expression for the dynamical aS Raman scattering cross-section:

$$\sigma_{aS}(\omega, t) = A V_{EM}^2 \, \rho_{phot}(\omega) \, J_E(\omega, t), \tag{S5}$$

where $J_E(\omega, t)$ denotes the *joint density of electronic states* that are available to contribute to the aS emission at the considered frequency ω and time delay $t$, while $A$ is a constant to be determined by comparison with the experiments. More precisely, similarly to the formalism introduced in Ref. 13, we defined $J_E(\omega, t)$ as follows:

$$J_E(\omega, t)=\int_{E_{min}}^{E_{max}} \rho_J(E, E+\hbar\omega, t)\mathrm{d}E, \tag{S6}$$

where $\rho_J$ is the energy distribution of the joint density of electronic states between the initial, higher energy, level $E_i$, and the final, lower energy, level $E_f$, involved in the emission of an aS photon of energy $\hbar\omega$ (namely such that $E_i = E_f + \hbar\omega$, with $E_f = E$ in Eq. S6 above). The expression for $\rho_J$ reads:

$$\rho_J(E_f, E_i, t) = \left[f(E_i, t)\, \rho_E(E_i)\right] \left\{\left[1 - f(E_f, t)\right] \rho_E(E_f)\right\}. \tag{S7}$$

In the above formula, $f(E, t)$ is the time-dependent nonequilibrium electron occupancy distribution (see Sections V and VII for details on how we calculated it for the photoexcitation conditions under analysis), and $\rho_E(E)$ is the electronic density of states, that we took from ab-initio calculations previously reported for Au[14]. For the extrema of integration in Eq. S6, we assumed $E_{max}$= 4.5 eV (from the Fermi level) and $E_{min} = -E_{max}$ thus ensuring to cover with numerical accuracy the entire energy spectrum of the non-equilibrium electronic distribution, featuring occupied states at energies as high as the probe photon energy (~1.7 eV).

With the set of expressions given above, our model provided us with a semi-quantitative estimation of the optical signal measured experimentally. Moreover, by considering the range of incident photon energies analysed in the experiments (below the Au interband transition threshold[15]), the description we pursued suggested that the measured aS emission can be rationalised as an electronic resonance Raman scattering mechanism.

**V. Modelling hot carrier ultrafast dynamics**

The ultrafast dynamics of high-energy 'hot' carriers resulting from the photoexcitation of the plasmonic picocavity with femtosecond laser pulses has been described numerically by a model adapted from the well-established Three-Temperature Model (3TM)[16]. Generally, the 3TM is a semiclassical rate-equation model for the femtosecond to picosecond dynamics of out-of-equilibrium plasmonic nanostructures[17], detailing the relaxation following ultrafast illumination in terms of three energetic variables: (i) $N$, the excess energy density stored in a fraction of non-equilibrium carriers referred to as 'nonthermal', as they feature an energy distribution differing by $\Delta f_{NT}$ from the equilibrium Fermi-Dirac distribution, characterised by a double-step-like spectral shape involving energies as high as the absorbed photon energy; (ii) $\Theta_E$, the temperature of the 'thermalized' portion of the hot carriers, whose energy distribution differs by $\Delta f_T$ from the equilibrium Fermi-Dirac distribution at room temperature $\Theta_0$, but can be described in terms of a Fermi-Dirac-like function at a higher temperature $\Theta_E$, i.e. $f(E,\Theta_E) = \left[1+e^{(E-E_F)/k_B\Theta_E}\right]^{-1}$; and (iii) $\Theta_L$, the lattice temperature of the nanostructure. By interlinking $N$, $\Theta_E$ and $\Theta_L$ in a set of three coupled ordinary differential equations, the 3TM accounts for the electron-electron and electron-phonon scattering mechanisms towards relaxation, and it thus allows for tracking over time the flow of the excess energy delivered by the incident photons to the metal electrons. To then retrieve the temporal evolution of the non-equilibrium (both nonthermal, NT, and thermal, T) carriers' population in terms of their differential energy occupancy distribution $\Delta f_{NT(T)}$, the following expressions apply, respectively:

$$\Delta f_{NT}(E, t) = \delta_{NT}(E)\, N(t), \qquad \text{(S8)}$$

$$\Delta f_{\mathrm{T}}(E,t) = f[E,\Theta_{\mathrm{E}}(t)] - f[E,\Theta_0], \tag{S9}$$

where $\delta_{\mathrm{NT}}(E)$ is a double-step-like function[16, 18] extending up to the incident photon energy from the Fermi level, and $f(E,\Theta_{\mathrm{E}})$ is the Fermi-Dirac distribution given above.

For the purpose of modelling the ultrafast response of the picocavity under analysis, given some peculiar features of such plasmonic architecture, we built upon the more conventional 3TM and developed an *ad-hoc* reduced rate-equation model for our system. In particular, dedicated simulations (exhaustively detailed in Section VI) showed that accounting for spatial diffusion of the thermalized hot carriers is critical to accurately model the system's ultrafast dynamics. As more extensively discussed below (refer to Section VI), this peculiar behaviour can be rationalised by considering the extreme localization of the optical modes across the picocavity, and the resulting strong spatial gradients of the electronic excitation. A rigorous description of these mechanisms can be achieved via an inhomogeneous formulation of the 3TM[19, 20], where three coupled partial differential equations regulate locally the space-time dynamics of hot carriers, yet at the cost of numerical complexity. Hence, to keep the model handy, we implemented a reduced approach and included the effects of the electronic spatial diffusion by introducing an imbalanced extra de-excitation channel for thermal carriers. The rate of this contribution was defined in terms of an effective diffusion time $\tau_{\mathrm{diff,eff}}$, and a polynomial in the electronic temperature was employed to mimic the actual diffusion process. Both the value of $\tau_{\mathrm{diff,eff}}$ and the polynomial order were then adjusted to best fit the rigorous solution obtained by solving (only once) the inhomogeneous 3TM (see Section VI). This reduced approach allowed us to drop the explicit space dependence in the model, and resort to a much simpler set of ordinary differential equations in the time-domain simulations.

In formulas, the reduced 3TM we implemented is given by the following set of coupled rate-equations:

$$\frac{\mathrm{d}N}{\mathrm{d}t} = -aN - bN + P_{\mathrm{abs,delay}}(t;t_{\mathrm{d}}), \tag{S10a}$$

$$C_{\mathrm{E}}\frac{\mathrm{d}\Theta_{\mathrm{E}}}{\mathrm{d}t} = aN - G(\Theta_{\mathrm{E}} - \Theta_{\mathrm{L}}) - \frac{1}{\tau_{\mathrm{diff,eff}}}(\Theta_{\mathrm{E}}^{\alpha} - \Theta_0^{\alpha}), \tag{S10b}$$

$$C_{\mathrm{L}}\frac{\mathrm{d}\Theta_{\mathrm{L}}}{\mathrm{d}t} = bN + G(\Theta_{\mathrm{E}} - \Theta_{\mathrm{L}}). \tag{S10c}$$

Here, in essence, $a = (h\nu_{\mathrm{p}})^2/(2\tau_0 E_{\mathrm{F}}^2)$ is the electron gas heating rate regulating the energy exchange from nonthermal to thermal electrons (with $h\nu_{\mathrm{p}}$ the incident photon energy, $E_{\mathrm{F}}$ the Au Fermi energy, and $\tau_0$ a material's constant, here set to 6 fs), $b(G)$ is the scattering rate from nonthermal (thermal) carriers to the phonons in the metal, $C_{\mathrm{E}} = \gamma_{\mathrm{E}}\Theta_{\mathrm{E}}$ ($C_{\mathrm{L}}$) is the electrons (lattice) heat capacity (with $\gamma_{\mathrm{E}}$ a constant), $\Theta_0$ is the temperature of the environment. Given the rather short (< 2 ps) timescales of our analysis, phonon-phonon coupling with the environment was not included in Eq. S10c. Further details on the derivation and values of these parameters can be found elsewhere[17]. Finally, the source term $P_{abs,delay}(t;\,t_d)$ in Eq. S10a denotes the pulse power density (per unit volume) absorbed by the plasmonic nanostructure, and can be expressed as:

$$P_{abs,delay}(t;\,t_d)=\sqrt{4\ln 2/\pi}\,\frac{\sigma_{abs}(\lambda_p)F_p}{V_{EM}\Delta t_p}\exp\left[-4\ln 2\,(t\,-\,t_d)^2/\Delta t_p^2\right], \tag{S11}$$

with $\sigma_{abs}$ and $V_{EM}$ the absorption cross-section and relevant optical mode volume introduced before, $F_p$ the pulse fluence, $\lambda_p$ its wavelength, and $\Delta t_p$ its temporal duration (full width at half maximum). Finally, $t_d$ is a delay time defining the centre of the Gaussian temporal envelope of the pulse, and acts as a fixed parameter to sweep over to model the pump-probe experiments.

## VI. Spatiotemporal diffusion of hot carriers

Considering the peculiar sub-nm confinement of the picocavity optical modes, the plasmonic hot carriers resulting from photoabsorption inherit a comparable extreme localization within the metal regions in close proximity to the gap. Strong spatial gradients are thus expected for the electronic excitation, with a relevant role of thermal electrons' ultrafast diffusion away from these intense-field regions. To substantiate this argument and quantify the impact of the electronic space-time evolution, we resorted to a fully inhomogeneous 3TM (I3TM)[19, 20] to describe the non-equilibrium dynamics of photoinduced carriers. Note that, in general, also nonthermal hot electrons could experience spatial diffusion[21]. However, we here neglect this effect since it would occur on timescales comparable with their relaxation[19], a process which is accomplished in a few hundreds of femtoseconds (mostly via electron-electron scattering) and expected to dominate their overall dynamics.

In brief, the I3TM extends the formulation of the original 3TM and, while keeping the same thermodynamical approach based on rate equations, it includes the space degree of freedom. Fourier-like terms have been proposed to describe the diffusion of thermal carriers and phonons, and a space-time source (function of **r** and *t*) is considered, based on the specific pattern of light absorption within the nanostructure. In formulas, following previous reports[19, 20], the partial differential equations constituting the model read as follows:

$$\frac{\partial N(\mathbf{r},t)}{\partial t} = -\,aN\,-\,bN\,+\,\rho_{abs}(\mathbf{r})P_{abs,delay}(t;\,t_d), \tag{S12a}$$

$$C_E\frac{\partial \Theta_E(\mathbf{r},t)}{\partial t} = -\nabla\cdot(-\kappa_E\,\nabla\Theta_E)\,-\,G(\Theta_E\,-\,\Theta_L)+aN, \tag{S12b}$$

$$C_L \frac{\partial \Theta_L(\mathbf{r}, t)}{\partial t} = \kappa_L \nabla^2 \Theta_L + G(\Theta_E - \Theta_L) + bN, \tag{S12c}$$

where the symbols introduced in Section V have the same meaning, $\kappa_E$ ($\kappa_L$) is the electron (lattice) thermal conductivity, and the quantity $\rho_{abs}(\mathbf{r})$, which encodes the space dependence of the drive term, is the (normalized) spatial pattern of the electromagnetic power dissipation density (refer to Section III). More details and the values of the parameters used can be found in previous works.[19, 20]

A 3D FEM-based model has been developed using COMSOL Multiphysics (version 6.2) to integrate numerically the I3TM of Eqs. S12a-12c. In the simulations, a simplified geometry (sketched in Fig. S7) was considered to reduce the computational cost, by leveraging that: (i) the spatial distribution of the electromagnetic absorption $\rho_{abs}(\mathbf{r})$ is strongly confined in a small volume around the gap between the STM tip and metal flat surface, so that regions far from the picocavity do not absorb almost any light, hence do not contribute to the generation of hot carriers; (ii) considering the electronic and the lattice temperatures, both the metal flat surface and the STM tip can be essentially treated as infinite sinks, given their macroscopic spatial extent compared to the nanoscale volume of the cavity mode.

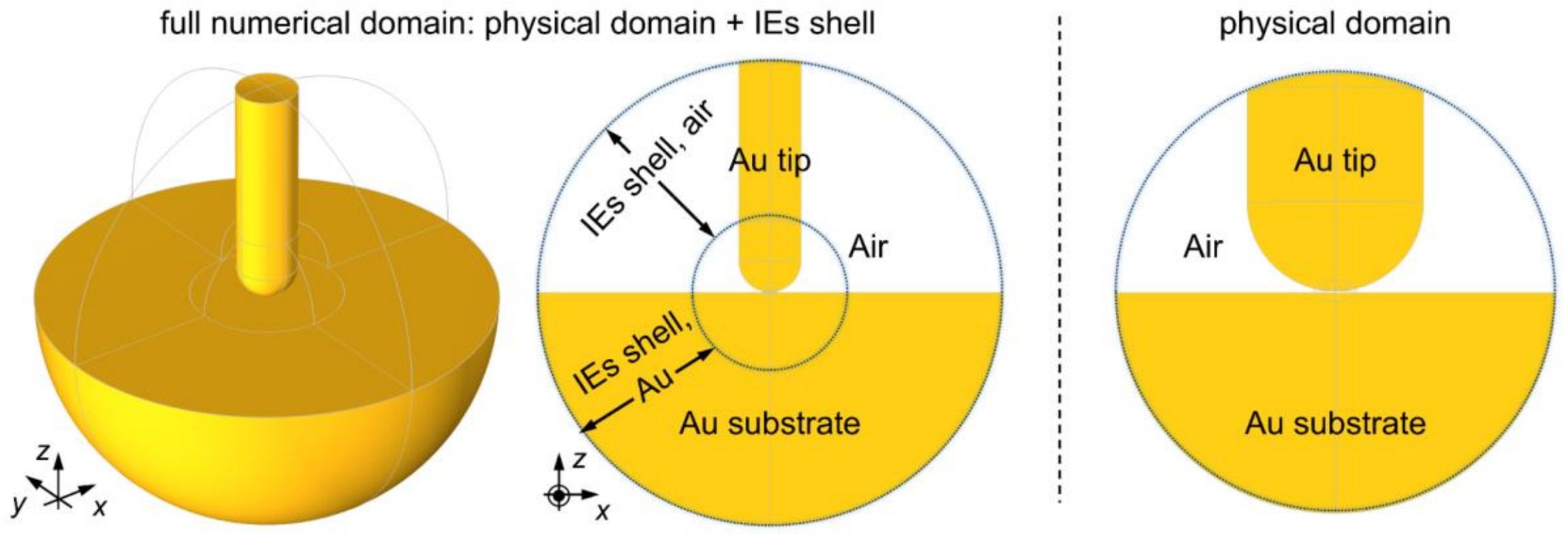

**Fig. S7 | Model for the spatiotemporal diffusion of hot carriers.** Simplified geometry used to solve numerically the I3TM. 3D (left) and lateral (middle) views of the full numerical domain, including the physical domain and a shell of Infinite Elements (IEs). In the physical domain (right), a sub-nm gap (0.5 nm) between a flat Au surface and a Au tip is considered. The spherical symmetry of the whole domain is chosen to reduce the computational burden of the 3D simulations.

For these reasons, we could reduce the physical domain used in the calculations to a smaller region close to the gap (defined by a radius of ~160 nm), and assign the COMSOL built-in Infinite Elements (IEs) to the surrounding domains, so as to mimic a semi-infinite extent for the air environment, the metallic tip, and the Au substrate. Prior to integrating the I3TM in the time domain, a stationary electromagnetic study was conducted (as detailed in Section III), to numerically estimate $\rho_{abs}(\mathbf{r})$ at the wavelength of the incident pulse and inherit it in Eq. S12a in a fully consistent manner. Finally, to accurately reproduce the experimental conditions, the temperatures initial values were set to $\Theta_E = \Theta_L = \Theta_0 = 11$ K.

As mentioned above (see also Section V), solving the I3TM served us specifically to: (i) corroborate the argument that carriers tend to move away from the proximity of the cavity on ultrafast timescales; and (ii) formulate a homogeneous, reduced rate-equation model (Eqs. S10a-S10c). We therefore tracked the spatiotemporal ultrafast dynamics of thermalized carriers, and the flow of their excess energy content across the metallic regions close to the gap between the tip and the substrate. For that, we monitored the evolution of the electronic temperature increase $\Delta\Theta_E(\mathbf{r},t)$, and the energy density stored in this carrier population fraction, expressed as:

$$\Delta E_E(\mathbf{r}, t) = \int_{-\infty}^{t} C_E(\mathbf{r}, t') \frac{\partial \Theta_E(\mathbf{r}, t')}{\partial t'} \mathrm{d}t'. \tag{S13}$$

In addition, to directly compare the outcome of the I3TM with homogeneous models, these two quantities were averaged over the optical mode volume $V_{EM}$, that, after inspecting $\rho_{abs}(\mathbf{r})$, we approximated by an ellipsoid of semiaxes ~11 nm ×11 nm × 4 nm, so as to obtain space-independent relevant quantities. To further facilitate the direct assessment of the carrier diffusion, calculations were performed in the absence of thermal electron-phonon coupling (i.e., $G = 0$ in Eqs. S12b, S12c), so that Fourier-like diffusion was the only available de-excitation channel for $\Theta_E$.

The main results of the simulations are summarised in Fig. S8, considering an incident pulse of 850 nm wavelength, 30 fs duration, ~0.15 $\mu$J/cm$^2$ fluence (i.e., typical experimental conditions). In particular, Fig.S8a shows a spatial map (2D cross-sectional view of the whole 3D geometry) of the electronic temperature field within the metallic regions, evaluated at 50 fs after the pulse peak, namely when $\Theta_E$ reaches its peak. The calculations reveal that a maximum temperature increase as low as ~30 K is obtained, which is remarkably low considering that, for the same excitation conditions, a homogeneous 3TM predicts a temperature increase of >500 K. Moreover, $\Delta\Theta_E(\mathbf{r},t)$ exhibits a strongly inhomogeneous spatial distribution, reminiscent of the optical cavity mode regulating the photo-absorption, with the largest values localized around the gap, and a quick drop when moving away from the picocavity region.

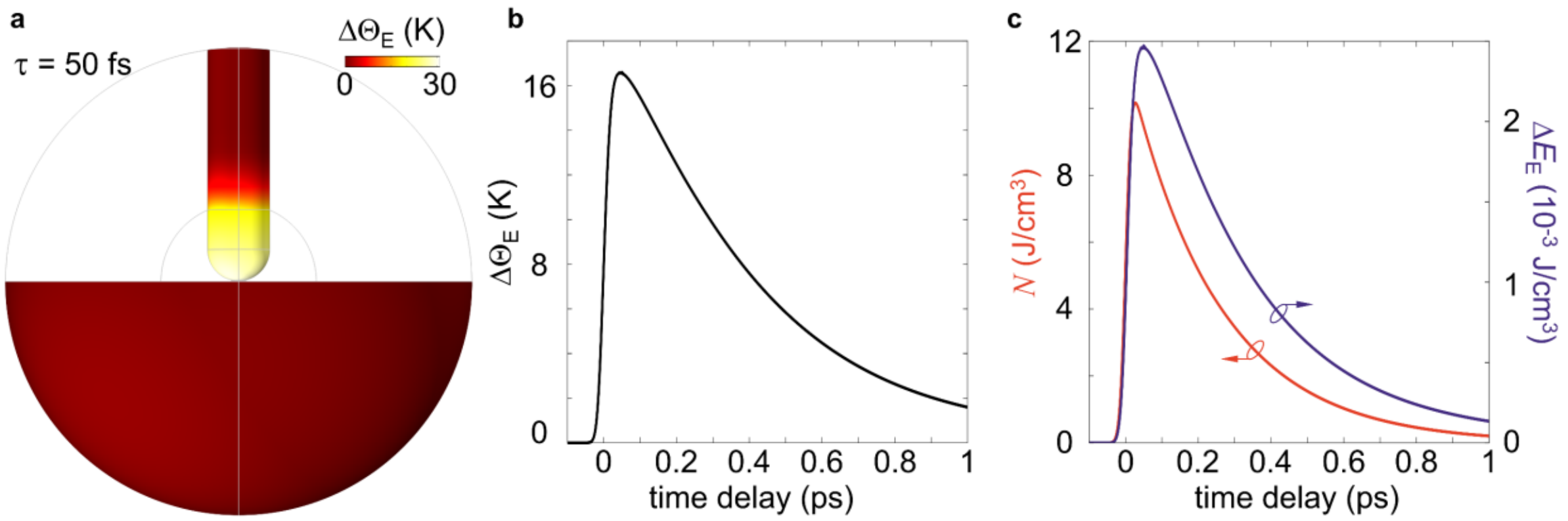

**Fig. S8 | Inhomogeneous spatiotemporal diffusion of hot carriers. a,** Spatial map (2D cross-sectional view) of the electronic temperature distribution within the metallic (tip and substrate) regions of the picocavity, evaluated at 50 fs after the pulse arrival. **b,** Ultrafast dynamics of $\Theta_E$, averaged over the mode volume. **c,** The ultrafast dynamics of the excess energy stored in the nonthermal portion of electrons $N$ (red, left vertical axis) is compared to that of the thermalized electron energy content $\Delta E_E$ (blue, right vertical axis). Both quantities are space-averaged over the cavity mode volume.

Due to such a strong confinement in very small volumes, extreme temperature gradients are obtained, making the diffusion term in Eq. S12b dominate over the carrier ultrafast dynamics. As displayed in Fig. S8b, $\Delta\Theta_E$ (averaged over the mode volume) undergoes a peculiar temporal evolution, and reaches almost full relaxation within ~1.5 ps, a noticeably short time compared to the standard dynamics of carriers in plasmonic nanostructures, requiring rather tens of ps to go back to equilibrium[17]. The same peculiar dynamics is observed also for $\Delta E_E$, which is reported in Fig. S8c and compared with the excess energy density stored in the nonthermal fraction of hot carriers, $N$ (both space-averaged over $V_{EM}$). Unlike any other more conventional plasmonic nanostructure, the two energy contents follow a comparable, sub-ps relaxation, with the contribution of thermal carriers being substantially smaller than the nonthermal ones (note the two distinct vertical axes of the plot, differing by ~3 orders of magnitude).

Finally, as the last step of our modelling approach, we leveraged the accurate solution of the I3TM to approximate the spatiotemporal diffusion of thermal electrons with an effective, more handy term to include in a space-independent rate equation for $\Theta_E$. We assumed this extra de-excitation term to be a polynomial (of order $\alpha$) in the electronic temperature associated with a characteristic time $\tau_{\mathrm{diff,eff}}$, namely of the form $-\frac{1}{\tau_{\mathrm{diff,eff}}}(\Theta_E^{\alpha} - \Theta_0^{\alpha})$, with $\Theta_0$ the environment temperature (here set to 11 K).

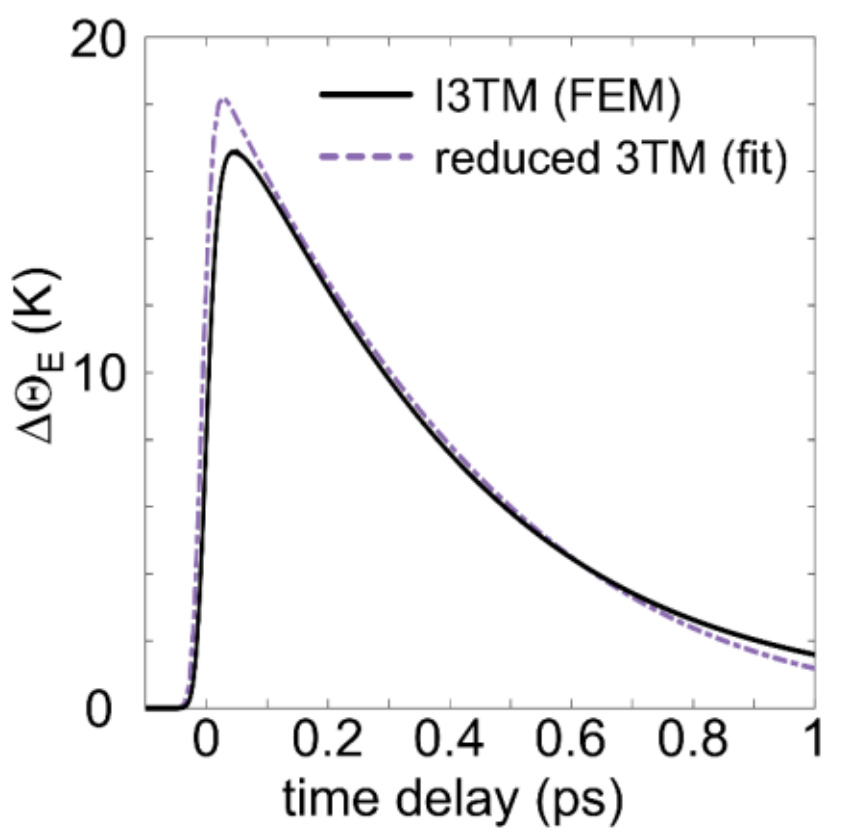

**Fig. S9 | Fit of the electronic temperature dynamics.** The ultrafast dynamics of the electronic temperature (averaged over the mode volume) obtained by integrating the I3TM in our FEM model (black solid line) is compared with that retrieved from a modified homogeneous 3TM (purple dashed line), where a polynomial of order $\alpha$ with a characteristic time $\tau_{diff,eff}$ mimics the electron temperature diffusion. The best fit is achieved for $\alpha = 3$ and $\tau_{diff,eff} = 0.7$ fs.

Both $\tau_{diff,eff}$ and $\alpha$ were used as fitting parameters. To determine their values, the dynamics of the space-averaged $\Delta\Theta_E$ obtained by integrating the I3TM in COMSOL (Fig. S8b) was compared with the numerical solution of the modified 3TM given by Eqs. S10a-S10c above. The best fit (results shown in Fig. S9) was found for a third-order polynomial ($\alpha = 3$ in Eq. S10b) and an effective diffusion time constant $\tau_{diff,eff} = 0.7$ fs. Note that the values obtained for these two parameters of the reduced 3TM should be intended as the result of a fitting procedure. That is, they represent effective quantities introduced to mimic the spatial diffusion of carriers in a simplified formulation of the electron spatiotemporal dynamics, hence do not immediately relate to the mechanisms behind of spatial diffusion. Their values are rather tuned to match the exact results obtained from the I3TM (Eqs. S12a-S12c), where spatial gradients and the electronic thermal conductivity regulate rigorously the evolution of carriers within the metal.

### VII. Modelling the ultrafast modulation of anti-Stokes emission

In order to simulate the pump-driven ultrafast modulation of the probe-induced anti-Stokes (aS) emission, the modelling ingredients outlined above were combined to determine the variations over time of the aS emitted power $P_{aS}$ (see Section IV) due to a non-equilibrium electronic distribution $\Delta f$.

More in detail, a nested approach was implemented, according to the following schematic algorithm:

(i) Pump excitation: our modified 3TM (Eqs. S10a-S10c, with the source term given by Eq. S11) was solved for an ultrashort pulse mimicking the interaction with the pump beam, centred around a reference time $t_d = 0$. All the relevant parameters of the simulated pulse (duration, central wavelength, fluence) were set as in the experiments. This calculation provided us with the ultrafast dynamics of the three energetic variables, $N_{pu}(t)$, $\Theta_{E,pu}(t)$, and $\Theta_{L,pu}(t)$, triggered by the pump photons absorption. The simulation was performed over a timescale comprised between $t_{-\infty} = -0.4$ ps and $t_{\infty} = 1.5$ ps.

(ii) Probe excitations: the outcome of the pump pulse simulation was used to set the initial conditions in a series of 3TMs mimicking the interaction with the probe, and iteratively solved by scanning the pump-probe time delay. Practically, we fixed a time delay $t_{d,i}$ and expressed the source term $P_{abs,delay}(t;t_{d,i})$ accordingly (Eq. S11), based on the experimental optical parameters of the probe pulse. We then built our 3TM (Eqs. S10a-S10c) for three variables $N_{pr}(t)$, $\Theta_{E,pr}(t)$, and $\Theta_{L,pr}(t)$, whose initial condition was the solution of the pump-excitation model at the considered time delay $t_{d,i}$, namely $N_{pr}(t = t_{-\infty}) = N_{pu}(t = t_{d,i})$, $\Theta_{E,pr}(t = t_{-\infty}) = \Theta_{E,pu}(t = t_{d,i})$, $\Theta_{L,pr}(t = t_{-\infty}) = \Theta_{L,pu}(t = t_{d,i})$. Integrating the rate equations of the 3TM provided us with the whole temporal evolution (from $t_{-\infty}$ to $t_{\infty}$) of the three probe-quantities, of which we extracted only their value at the delay of interest, namely $N_{pr}(t = t_{d,i})$, $\Theta_{E,pr}(t = t_{d,i})$, and $\Theta_{L,pr}(t = t_{d,i})$. Then, to reconstruct the actual dynamics as a function of the pump-probe delay, the parameter $t_{d,i}$ was scanned from $t_{-\infty}$ to $t_{\infty}$, and the probe-3TM was iteratively solved for each value of the delay. By collecting the series of solutions at each $t_{d,i}$, we thus retrieved $N_{pr}(t_d)$, $\Theta_{E,pr}(t_d)$, and $\Theta_{L,pr}(t_d)$ over the entire pump-probe delay time window relevant to the experiments.

(iii) Electronic distribution and aS emission rate: we considered, in the most general conditions, the electronic contribution to the aS emission (and in particular, the aforementioned joint density of electronic states $J_E(\omega, t)$, Eq. S6 in Section IV) as the superposition of three terms related to (i) the equilibrium occupancy distribution of electrons, hence time-independent and typically very weak, especially at cryogenic temperature; (ii) the excitation driven by the pump, centred at time delay $t_d = 0$ and expressed by the carrier distribution modulations $\Delta f_{NT,pu}(E,t)$ and $\Delta f_{T,pu}(E,t)$ (refer to Eqs. S8 and S9, respectively), evolving in time according to the dynamics of $N_{pu}(t)$ and $\Theta_{E,pu}(t)$ obtained in the first step above; and (iii) the excitation driven by the probe, depending on the pump-probe time delay and mostly dominating over the others. To compute this term, we employed the dynamics of $N_{pr}(t_d)$ and $\Theta_{E,pr}(t_d)$ obtained from the nested 3TMs in the previous step, and determined the corresponding non-equilibrium distribution $f_{pr}(E,t_d) = f(E, \Theta_0) + \Delta f_{NT,pr}(E,t_d) + \Delta f_{T,pr}(E,t_d)$.

Finally, each of these terms was plugged in the (differential with respect to equilibrium) expressions of $\rho_J$ (Eq. S7) and $J_E(\omega, t)$ (Eq. S6) to calculate the aS power $P_{aS}$.

## VIII. Calculations of the picocavity LPDOS in the presence of a graphene nanoribbon (GNR)

Further numerical calculations were performed to complement the measurements conducted on the plasmonic picocavity when a graphene nanoribbon (GNR) is inserted into the sub-nm gap between the STM tip and the flat Au surface (refer to Fig. 4 in the main text). In particular, we examined the local photonic density of states (LPDOS) of the picocavity $\rho_{phot}$ in the presence of the GNR as a function of the nanoribbon position. A simulation equivalent to the one presented in Section II was implemented, allowing us to determine the system's electromagnetic response under dipolar excitation, and to calculate $\rho_{phot}$ (of which again we analysed specifically the radiative component) for the modified cavity configuration. The GNR was accounted for in the model by defining a rectangular 2D domain (sheet of zero thickness) of size 5 nm × 1 nm (consistent with the experimental sample) lying on the Au substrate, with an assigned conductivity $\sigma(\omega)$. The analytical expression of $\sigma(\omega)$ given in previous reports[22, 23] was considered, assuming typical values for the parameters of graphene that were not characterised experimentally. Specifically, we used a Fermi energy of 0.2 eV, and a Drude scattering time of 100 fs. The contribution of the GNR to the optical response of the whole system was then expressed in terms of an induced surface current $\boldsymbol{j} = \sigma\boldsymbol{E}$, with $\boldsymbol{E}$ the electric field solution of Maxwell's equations.

To investigate the impact of the relative position of the GNR on the photonic properties of the picocavity, calculations were conducted by varying gradually the in-plane position of the GNR with respect to the Au tip. Starting from the precise centre of the cavity, right below the tip ($x_{GNR} = 0$, $y_{GNR} = 0$, as sketched in Fig. S10a, top), the sheet was moved 2.5 nm apart in the $y$-direction (Fig. S10a, bottom), namely completely out of the cavity.

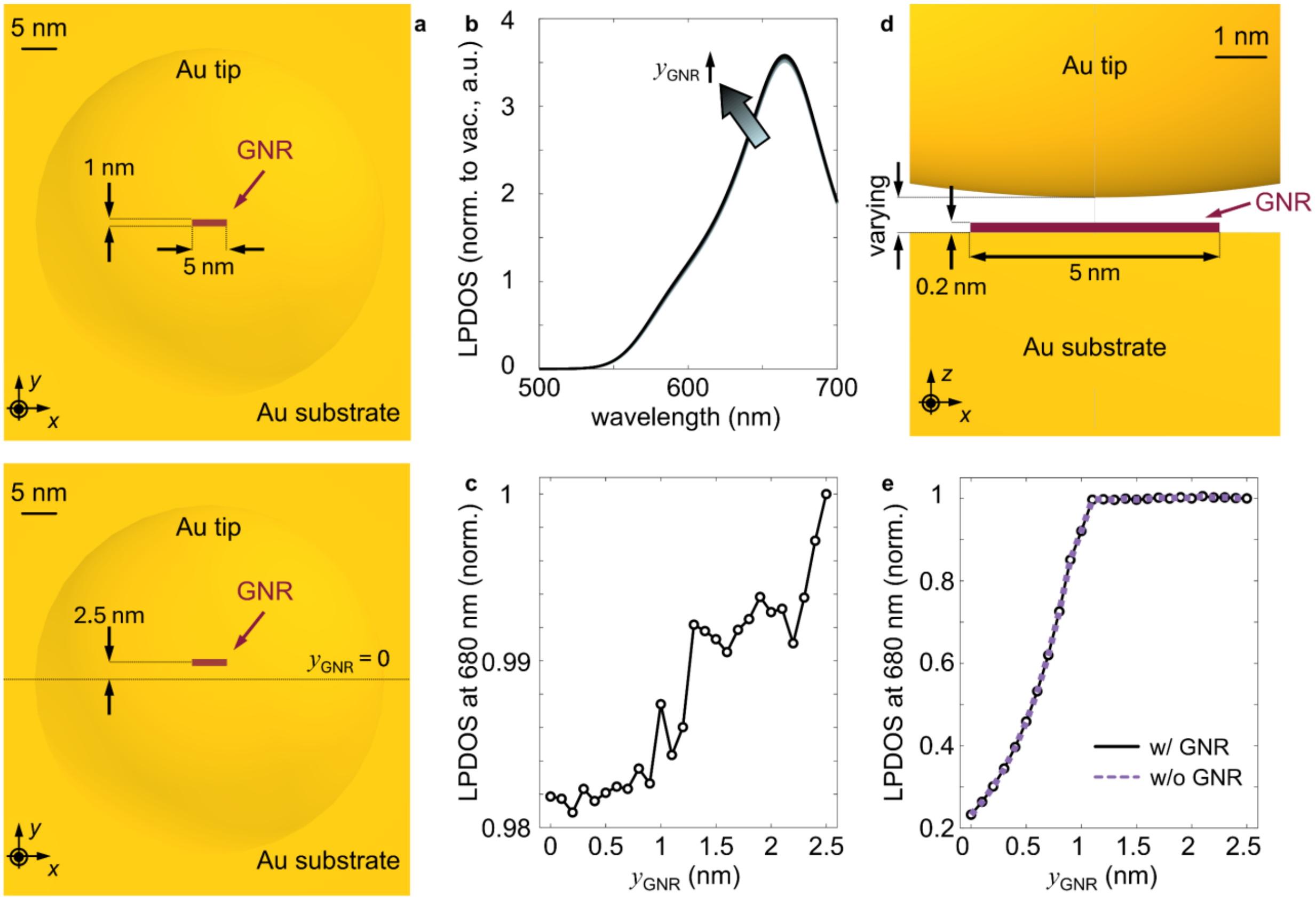

**Fig. S10 | LPDOS with GNR. a,** Top view of the cavity, where a sheet of GNR is inserted and moved from the centre (top) outside the cavity (bottom). **b,** LPDOS calculated at various positions of the GNR. **c,** LPDOS at 680 nm as a function of the *y*-coordinate of the GNR. **d,** Schematic of the 3D GNR model, where the tip-distance changes according to the in-plane position of the GNR. **e,** LPDOS at 680 nm calculated for the 3D GNR by concurrently moving the GNR in the plane and varying the distance between the tip and the Au substrate surface. Dashed line is the calculation when the GNR is not in place. The tip-substrate distance is varied linearly with the $y_{GNR}$ from 0.7 nm (for $y_{GNR}$ = 0) to 0.5 nm (when $y_{GNR}$ = 1.25 nm).

For each position of the nanoribbon, the corresponding LPDOS was computed, according to the formulas given in Section II. Despite the displacement of the graphene sheet, the calculated spectra of $\rho_{phot}$ did not exhibit any significant change. Our main results are summarised in Fig. S10b, where indeed all curves, each of which corresponds to a distinct $y_{GNR}$, mostly overlap. More specifically, in Fig. S10c we examined the evolution of the LPDOS at a wavelength of 680 nm, i.e., at which the aS signal assigned to hot carriers was recorded in the experiments (refer to Fig. 4 in the main text). Again, only minor changes were observed when moving the nanoribbon across the picocavity. In these terms, our numerical model did not suggest that a photonic Au-GNR interaction could cause substantial variations in the LPDOS.

In addition to this set of simulations, we conducted some complementary calculations of the same kind, where we yet described the GNR as a 3D layer of finite thickness $t_{GNR}$ = 0.2 nm, and same in-plane size (5 nm × 1 nm). The modelled geometry was modified accordingly (schematically depicted in Fig. S10d), and, in terms of graphene optical properties, we introduced[24, 25] an anisotropic dielectric permittivity with in-plane components $\varepsilon_{xx}(\omega) = \varepsilon_{yy}(\omega) = 1 - i\sigma(\omega)/(\omega\varepsilon_0 t_{GNR})$. Indeed, although the optical description of GNRs characterised by a conductivity $\sigma(\omega)$ is well-established, we introduced a finite thickness of the GNR to examine possible effects resulting from the changes in the cavity effective size. Considering that: (i) the GNR thickness $t_{GNR}$ estimated experimentally was comparable with the gap distance between the tip and the Au substrate; and (ii) the STM operated in constant current mode, thus the tip moved upward (from 0.5 nm up to ~ 0.65 nm from the Au surface) when sitting on top of the GNR, our additional simulations aimed at reproducing these spatial changes in the picocavity configuration, and assessing their impact. The calculations were therefore performed following the same procedure as the 2D-sheet case, but this time each value of $y_{GNR}$ corresponded to a distinct value of the distance between the tip and the Au substrate, starting from 0.65 nm for $y_{GNR}$ = 0, and back to 0.5 nm when $y_{GNR}$ = 1.25 nm (i.e., the entire 1-nm-wide GNR can be considered to be outside the cavity). The so-obtained LPDOS, evaluated at 680 nm, is reported in Fig. S10e (solid line), and seems to exhibit a variation (by almost a factor of 4) with increasing $y_{GNR}$. We argue however that the observed trend should be ascribed to the change in the size of the gap between the STM tip and the Au substrate, rather than to photonic effects introduced by the GNR. To ascertain this argument, we repeated the same calculation with the moving tip, yet by removing the GNR from the picocavity. The resulting $\rho_{phot}$, shown as well in Fig. S10e (dashed line), follows practically the same trend as the study with the GNR in place, indicating that the gap size is in fact responsible for the variations of LPDOS in our calculations.

The measured line profiles of the variation of the anti-Stokes signal intensity when the nanotip makes a lateral line scan across the GNR as shown in Fig. 4b has peculiar features at the edges of the nanoribbon (i.e. local maxima). These feature cannot be reproduced by the simulation of the LPDOS accounting only for the dielectric property of the GNR. Thus, these features should be originating from the contributions of the electronic structure of the GNR to the LPDOS (as also argued in the main-text), which were not considered in the simulations.